\documentclass[twocolumn,floatfix]{aastex631}
\usepackage{float}
\usepackage{graphicx}

\newcommand{\rev}[1]{#1}

\newcommand{\revv}[1]{#1}

\newcommand{\revt}[1]{#1}

\shorttitle{CHARIS HIP 93398B}
\shortauthors{Lewis}
\graphicspath{{./}{figures/}}

\begin{document}

\title{SCExAO/CHARIS Spectroscopic Characterization of Cloudy L/T Transition Companion Brown Dwarf HIP 93398 B}

\correspondingauthor{Briley Lewis}
\email{brileylewis@g.ucla.edu}

\author[0000-0002-8984-4319]{Briley L. Lewis}
\affiliation{Department of Physics and Astronomy, UCLA, Los Angeles, CA 90095 USA}

\author[0000-0002-6845-9702]{Yiting Li}
\affiliation{Department of Physics, UCSB, Santa Barbara, CA 93106 USA}

\author[0000-0002-9027-4456]{Aidan Gibbs}
\affiliation{Department of Physics and Astronomy, UCLA, Los Angeles, CA 90095 USA}

\author[0000-0002-0176-8973]{Michael P. Fitzgerald}
\affiliation{Department of Physics and Astronomy, UCLA, Los Angeles, CA 90095 USA}

\author[0000-0003-2630-8073]{Timothy Brandt}
\affiliation{Department of Physics, UCSB, Santa Barbara, CA 93106 USA}

\author[0000-0001-8170-7072]{Daniella Bardalez Gagliuffi}
\affiliation{Department of Physics and Astronomy, Amherst College, Amherst, MA 01002 USA}

\author[0000-0003-0115-547X]{Qier An}
\affiliation{Department of Physics, UCSB, Santa Barbara, CA 93106 USA}

\author[0000-0001-8892-4045]{Minghan Chen}
\affiliation{Department of Physics, UCSB, Santa Barbara, CA 93106 USA}

\author[0000-0001-5831-9530]{Rachel Bowens-Rubin}
\affiliation{Department of Astronomy, UCSC, Santa Cruz, CA 95064 USA}

\author[0000-0002-5082-6332]{Ma\"issa Salama}
\affiliation{Department of Astronomy, UCSC, Santa Cruz, CA 95064 USA}

\author[0000-0002-3047-1845]{Julien Lozi}
\affiliation{Subaru Telescope, National Astronomical Observatory of Japan, 650 N. Aohoku Pl., Hilo, HI 96720, USA}

\author[0000-0003-0054-2953]{Rebecca Jensen-Clem}
\affiliation{Department of Astronomy, UCSC, Santa Cruz, CA 95064 USA}

\author[0000-0003-0526-1114]{Ben Mazin}
\affiliation{Department of Physics, UCSB, Santa Barbara, CA 93106 USA}



\begin{abstract}

Brown dwarfs with measured dynamical masses and spectra from direct imaging are benchmarks that anchor substellar atmosphere cooling and evolution models. We present Subaru SCExAO/CHARIS infrared spectroscopy of HIP 93398 B, a brown dwarf companion recently discovered by \citet{Li2023MNRAS} as part of an informed survey using the Hipparcos-Gaia Catalog of Accelerations. This object was previously classified as a T6 dwarf based on its luminosity, with its independently-derived age and dynamical mass in tension with existing models of brown dwarf evolution. Spectral typing via \rev{empirical standard spectra}, temperatures derived by fitting substellar atmosphere models, and $J-H$, $J-K$ and $H-L'$ colors all suggest that this object \revv{has a substantially higher temperature and luminosity, consistent with classification as} a late-L dwarf near the L/T transition \revv{(T = 1200$^{+140}_{-119}$ K)} with moderate to thick clouds \rev{possibly} present in its atmosphere. When compared with the latest generation of evolution models that account for clouds with our revised luminosity and temperature for the object, the tension between the model-independent mass/age and model predictions is resolved.

\end{abstract}



\section{Introduction} \label{sec:intro}

Brown dwarfs are substellar mass objects that may burn lithium or deuterium ($\sim$13--$\sim$80 $M_{\rm Jup}$), but are not sufficiently massive to sustain hydrogen fusion and enter the main sequence as a star \citep{1963PThPh..30..460H,1963ApJ...137.1121K}. Over their lifetimes, they cool and evolve through spectral types late-M, L, T, and Y \citep{1999PASP..111.1591M, mason2006astrophysics}. Of particular interest is the L/T transition, a stage characterized by a major change in infrared color, but only a minor change in temperature; additionally, T dwarfs are characterized by methane absorption that is stronger in later-type objects \citep{burgasser2006hubble,helling2014atmospheres}. As substellar objects cool, their atmospheric chemistries evolves, and various species of clouds are able to condense \citep{lodders2006chemistry,kirkpatrick2005new}. Multiple generations of substellar atmosphere models of increasing complexity in their treatment of atmospheric composition, cloud formation, and vertical transport, have been generated to describe the cooling evolution of brown dwarfs across their NIR spectra\citep{1996Sci...272.1919M, 2002AandA...382..563B,2014ApJ...787...78M}, such as the Sonora series of model grids \citep{marley2021sonora,morley2024sonora,mukherjee2024sonora}. These models have also been used to estimate masses of directly-imaged substellar objects based on their observed luminosities and spectra. One of the key challenges of brown dwarf characterization is the degeneracy present between age, luminosity, and mass due to the way they cool over time. Systems with  independently-derived ages, masses, and spectral types are therefore key benchmarks for testing substellar atmosphere models, especially as these models increase in complexity, including factors such as disequilibrium chemistry and clouds \citep{marley2015cool,morley2024sonora,mukherjee2024sonora}.

Informed direct imaging searches using the Hipparcos-Gaia Catalog of Accelerations \citep{Brandt2018ApJS,Brandt2021ApJS} have recently yielded numerous discoveries of new brown dwarf companions \citep{2019MNRAS.490.1120F, 2022MNRAS.513.5588B,Franson2023AJ}, including some that are amenable for direct spectroscopy such as those observed with the SCExAO/CHARIS instrument at Subaru Observatory \citep{guyon2010subaru,groff2016laboratory,2022AJ....164..152S,2021SPIE11823E..04C,kuzuhara2022direct,chilcote2021scexao}. For systems with radial velocity measurements, it is also possible to determine a companion's dynamical mass, a measurement independent of models describing an object's luminosity evolution \citep{Brandt+Dupuy+Bowler_2019}. To date, around 30 brown dwarf companions have been imaged and characterized using these combined techniques \citep{Li2023MNRAS}. 

Modeling the atmospheres of L/T transition objects has been a challenge for decades; accurate models require thorough line lists for a variety of molecules, many of which have only recently become available, and a consideration of many parameters, particularly gravity and metallicity, to which resulting spectra are quite sensitive \citep{morley2024sonora}. Additionally, the treatment of clouds has been a major issue facing substellar atmosphere models, especially in finding self-consistent evolutionary and spectral models. Until recently, many models assumed cloud-free atmospheres, despite the fact that disequilibrium chemistry and clouds are known to be important factors in brown dwarf atmospheres, especially for L/T transition objects \citep{marley2015cool}. 

Many brown dwarfs do agree with evolutionary models, supporting the models' validity (e.g. $\varepsilon$ Indi BC; \citealt{chen2022precise}). However, a few recently discovered T dwarfs show a slight tension with models (e.g. \citet{1995Sci...270.1478O}, \citet{2018AandA...614A..16C}, \citet{2021ApJ...913L..26B}). It has been suggested that this tension may be a result of an unresolved additional companion, contributing to the measured mass but not substantially to the luminosity and observed spectrum. Alternatively, this tension may be due to missing physics in atmosphere models.  

The massive companion brown dwarf HIP 93398 B was initially discovered with Keck/NIRC2 by \citet{Li2023MNRAS} as one of the first objects whose location was predicted from radial velocity and astrometric accelerations prior to imaging, and suggested as another T dwarf in tension with evolution models. Here, we present follow-up imaging and spectroscopy of the object using Subaru SCExAO/CHARIS \citep{guyon2010subaru,groff2016laboratory}. We aim to more confidently establish this object's \revv{temperature, luminosity, and} spectral type, determining whether or not it is in tension with brown dwarf cooling models as previously suggested. With a well-constrained dynamical mass and known spectral type, HIP 93398 B has the potential to serve as a benchmark brown dwarf for testing substellar atmosphere models.

In this work, we begin with an overview of the system's properties, including previously published information on the companion HIP 93398 B from the recent initial discovery by \citet{Li2023MNRAS}, in Section \ref{sec:properties}. We describe our new observations of the object using Subaru SCExAO/CHARIS in Section \ref{sec:data}. Section \ref{sec:orbit} presents an update to the companion's orbital fit, and Section \ref{sec:spec} presents spectral typing and photometry of HIP 93398 B, as well as a comparison to the newly updated Sonora Diamondback \citep{morley2024sonora} substellar atmosphere models, which include both evolutionary and spectral models with clouds for warm L and early T dwarfs. Finally, we discuss the implications of these findings for the ``overmassive T dwarf'' problem and conclude in Section \ref{sec:disc}.

\section{System Properties}\label{sec:properties}

HIP 93398 (HD 176535) is a main-sequence K3.5V star \citep{Gondoin2020AandA} at a distance of 36.99$\pm$0.03 parsecs, as measured by \textit{Gaia} EDR3 \citep{GaiaCollaboration2022AandA}. Many of its properties (summarized in Table \ref{tab:sysprops}) are similar to the Sun's, and it has been identified as a possible ``sibling'' from the same birth cluster \citep{Batista2014AandA,Adibekyan2018AandA}. This star has multiple age estimates with large uncertainties, but no conclusive precise value. Measurements of the chromospheric index $\log R'_{\rm HK} = -4.732$ indicate that this star is old and inactive, with a weakly constrained age of $5.57 \pm 4.84$ Gyr \citep{GomesdaSilva2021AandA}. The system's isochronal age, based on a sample of FGK stars from the HARPS sample with \textit{Gaia} parallaxes and using the PAdova and TRieste Stellar Evolution Code (PARSEC) \citep{Bressan2012MNRAS}, is 5.21$\pm$4.68 Gyr \citep{DelgadoMena2019AandA}. Meanwhile, the \textit{Gaia} collaboration measures an isochronal age of 3.34$^{+6.97}_{-0.61}$ Gyr with the PARSEC isochrones and their Fitting Location of Age and Mass with Evolution (FLAME) model \citep{GaiaCollaboration2022AandA}. Recent work from \citet{Li2023MNRAS} uses S-index values \revv{(a proxy for stellar magnetic activity)} of 0.43 \citep{GomesdaSilva2021AandA} and 0.63 \citep{DelgadoMena2019AandA} and Bayesian-based method for age estimation developed by \citet{Brandt2014ApJ} to estimate an age of 3.59$^{+0.87}_{-1.15}$ Gyr. In this paper, we adopt an age assumption of 3.59$^{+0.87}_{-1.15}$ Gyr as in \citet{Li2023MNRAS}.

The recently discovered brown dwarf companion HIP 93398 B was inferred to be a T dwarf based on its age, mass, and luminosity \citep{Li2023MNRAS}. According to that work, the companion is on a 41.3$^{+3.6}_{-3.1}$ year orbit, with a measured dynamical mass of $65.9^{+2.0}_{-1.7}\,M_{\rm Jup}$. The companion's other known properties are also listed in Table \ref{tab:sysprops}. This was one of the first companions with an astrometric location known prior to direct imaging. \rev{However, based on its derived bolometric luminosity (obtained by estimating W1 from L' and using relationships between W1 and bolometric luminosity), it is somewhat overmassive compared to expectations from evolutionary models}, possibly indicating a tension between models and observations or that the companion is actually an unresolved binary \citep{Li2023MNRAS}.

\begin{deluxetable}{lcr}
\tablecaption{Properties of the HIP~93398 System \label{tab:sysprops}
}
\tablewidth{0pt}
\tablehead{
Property & Value & References\tablenotemark{a}
}
\startdata
\multicolumn{3}{c}{Host Star (A)} \\ \hline
$\varpi$ (mas) & $27.033 \pm 0.018$ & 1 \\
Distance (pc) & $36.99 \pm 0.03$ & 1 \\ Spectral Type & K3.5V & 2,3 \\ 
Mass (M$_\odot$) & $0.72 \pm 0.02$ & 4,5 \\ 
Age (Gyr) & $3.59^{+0.87}_{-1.15}$ & 12 \\ 
$T_{\rm eff}$ (K) & $4727 \pm 104$ & 6 \\ 
$[{\rm Fe/H}]$ (dex) & $-0.15 \pm 0.07$ & 7 \\ 
$\log R'_{\rm HK}$ (dex) & $-4.85$ & 8 \\ 
$\log R_X$ (dex) & $<$$-4.28$ & 9 \\ 
$\log g$ (dex) & $4.64 \pm 0.10$ & 14 \\ 
\textit{Gaia} RUWE & 1.019 & 1 \\ 
Luminosity (L$_\odot$) & $0.208 \pm 0.007$ & 12 \\ 
\textit{Gaia} G (mag) & $9.374 \pm 0.003$ & 1 \\ 
$B_T$ (mag) & $11.195 \pm 0.066$ & 10 \\ 
$V_T$ (mag) & $9.923 \pm 0.035$ & 10 \\ 
$J$ (mag) & $7.804 \pm 0.027$ & 11 \\ 
$H$ (mag) & $7.313 \pm 0.033$ & 11 \\ 
$K_s$ (mag) & $7.175 \pm 0.020$ & 11 \\\hline 
\multicolumn{3}{c}{Companion (B)} \\ \hline
$H$ apparent (mag) & $16.51 \pm 0.57$ & 13 \\ 
$J$ apparent (mag) & $17.54 \pm 0.39$ & 13 \\ 
$K$ apparent (mag) & $15.96 \pm 0.13$ & 13 \\ 
$L'$ apparent (mag) & $16.31 \pm 0.07$ & 12 \\ 
Mass ($M_{\rm Jup}$) & $65.9^{+2.0}_{-1.7}$ & 12 \\
Semimajor axis (au) & $11.05^{+0.64}_{-0.56}$ & 12 \\ 
Inclination ($^\circ$) & $48.9^{+3.4}_{-3.7}$ & 12 \\ 
Period (yr) & $40.6^{+3.9}_{-3.5}$ & 12 \\ 
Eccentricity & $0.496^{+0.022}_{-0.020}$ & 12 \\ \hline
\multicolumn{3}{c}{\revt{Companion (B) from Li et al. 2023}} \\ \hline
\revt{Spectral Type} & \revt{T6.0} & \revt{12}\\
\revt{Temperature (K)} & \revt{$980\pm35$} & \revt{12}\\
\revt{$\log(L_{\rm bol}/L_\odot)$} & \revt{$-5.26\pm0.07$} & \revt{12} \\ \hline
\multicolumn{3}{c}{\revt{Companion (B) from This Work}} \\ \hline 
\revt{Spectral Type} & \revt{L9.0} & \revt{13} \\ 
\revt{Temperature (K)} & \revt{$1200^{+140}_{-119}$} & \revt{13} \\ 
\revt{$\log(L_{\rm bol}/L_\odot)$} & \revt{\rev{$-4.64 \pm 0.23$}} & \revt{13} \\ \hline
\enddata
\tablenotetext{a}{References abbreviated as (1) \citet{GaiaCollaboration2020AandA}; (2) \citet{Gray2006AJ}; (3) \citet{Bourges2014ASPC}; (4) \citet{Reiners2020ApJS}; (5) \citet{DelgadoMena2019AandA}; (6) \citet{Sousa2011AandA}; (7) \citet{Gaspar2016ApJ}; (8) \citet{Pace2013AandA}; (9) \citet{Voges1999ICRC}; (10) \citet{Hog2000eaa}; (11) \citet{Cutri2003yCat}; (12) \citet{Li2023MNRAS}; (13) this work; and (14) \citet{Stassun_TIC_2019}.}
\end{deluxetable}

\section{Observations and Data Reduction}\label{sec:data}

The HD~176535 system was previously observed with the ESO HARPS spectrograph \citep{Mayor2003Msngr, Trifonov2020AandA} for radial velocity measurements, and its astrometric acceleration was identified via \textit{Hipparcos} and \textit{Gaia} data in the Hipparcos-Gaia Catalog of Accelerations \citep{Brandt2018ApJS,Brandt2021ApJS}, allowing for the prediction of the astrometric location of the B companion in \citet{Li2023MNRAS}. \citet{Li2023MNRAS} also obtained $L'$ band Keck/NIRC2 AO imaging data of the system, revealing the companion at the predicted astrometric location. In this work, we add follow-up imaging spectroscopy using the Coronagraphic High Angular Resolution Imaging Spectrograph \citep[CHARIS,][]{groff2016laboratory} behind the Subaru Coronagraphic Extreme Adaptive Optics \citep[SCExAO,][]{guyon2010subaru,Jovanovic2015PASP} system on the 8.2-meter Subaru Telescope on Maunakea in Hawai'i.

\subsection{SCExAO/CHARIS High-Contrast Imaging \& Spectroscopy}\label{subsec:obs}

We observed HIP 93398 on 2022 October 7 UTC in CHARIS broadband mode (simultaneous observations in $J$, $H$, and $K$ bands -- 1.16-2.37 $\mu$m with R$\sim$18 \citep{groff2016laboratory} and high-resolution H-band (centered on 1.63 $\mu$m, $R\approx 65$). We additionally acquired high-resolution $J$-band data (centered on 1.24 $\mu$m with $R \approx 75$) on 2023 September 1 UTC. 
Seeing conditions were a steady $\sim$0$.\!\!''5$ across the 2022 night and were variable between $\sim$0$.\!\!''5$-0$.\!\!''8$ on the 2023 night
, according to the MASS/DIMM data available from the CFHT seeing and mass profile records. The individual exposure time was 60 seconds for broadband, 40 seconds for \textit{J} band, 
and 60 seconds for \textit{H} band, with total on-source integration times of 34 minutes, 80 minutes, and 82 minutes. All observations were taken with the instrument's 139 mas coronagraph. Observations and measured companion characteristics are summarized in Table \ref{tab:obs}.

\begin{deluxetable}{lccr}
\tablewidth{0pt}
\tablecaption{SCExAO/CHARIS observations and relative astrometry of HIP~93398~AB \label{tab:obs}}
\tablehead{
Filter &
Date (UT) &
Sep.~(mas)\tablenotemark{a} &
PA ($^\circ$)\tablenotemark{a}
}
\startdata
Broadband & 2022-10-07 & $368 \pm 11$ & $96.7 \pm 1.0$ \\
$H$ & 2022-10-07 & $370 \pm 11$ & $96.7 \pm 1.0$ \\
$J$ & 2023-09-01 & $387 \pm 11$ & $99.9 \pm 1.0$ 
\enddata
\tablenotetext{a}{Similar to \citet{Li2023MNRAS}, we adopt 3\% fractional errors for separation and 3\% for position angle.}
\end{deluxetable}

For these observations, an initial wavefront correction was applied by AO188 \citep{Hayano2008SPIE}, followed by higher-order corrections from SCExAO \citep{guyon2010subaru,Jovanovic2015PASP}. A Lyot coronagraph  within SCExAO with 3$\lambda$/D inner working angle then masks the central star for observation with the CHARIS integral field spectrograph (IFS). The IFS allows for spectral differential imaging (SDI; \citealt{Oppenheimer2009ARAandA}), and the observations were performed using SCExAO's 
pupil-tracking mode to enable angular differential imaging (ADI; \citealt{Marois2006ApJ}). All observations also included ``satellite spots'' created by modulations imposed on the SCExAO deformable mirror for astrometric and spectrophotometric calibration, using a 15$\lambda / D$ separation and 0.05 $\mu$m amplitude \citep{Jovanovic_Speckles2015,sahoo2020precision}.

\subsection{Data Reduction with CHARIS-DEP \& pyklip-CHARIS}

Data cubes (with wavelength as the third dimension) were extracted from the raw CHARIS data and calibrated using the instrument's standard pipeline (\texttt{CHARIS-DEP}) from \citet{Brandt2017JATIS}, which includes crosstalk corrections and removal of correlated noise. We then performed PSF subtraction using \texttt{pyklip-CHARIS} \citep{Chen2023RASTI}\footnote{Detailed information and tutorials on CHARIS-specific additions to \texttt{pyklip} can be found at \url{https://pyklip.readthedocs.io/en/latest/instruments/CHARIS.html}}. 

\begin{figure*}
    \centering
    \includegraphics[width=0.75\linewidth]{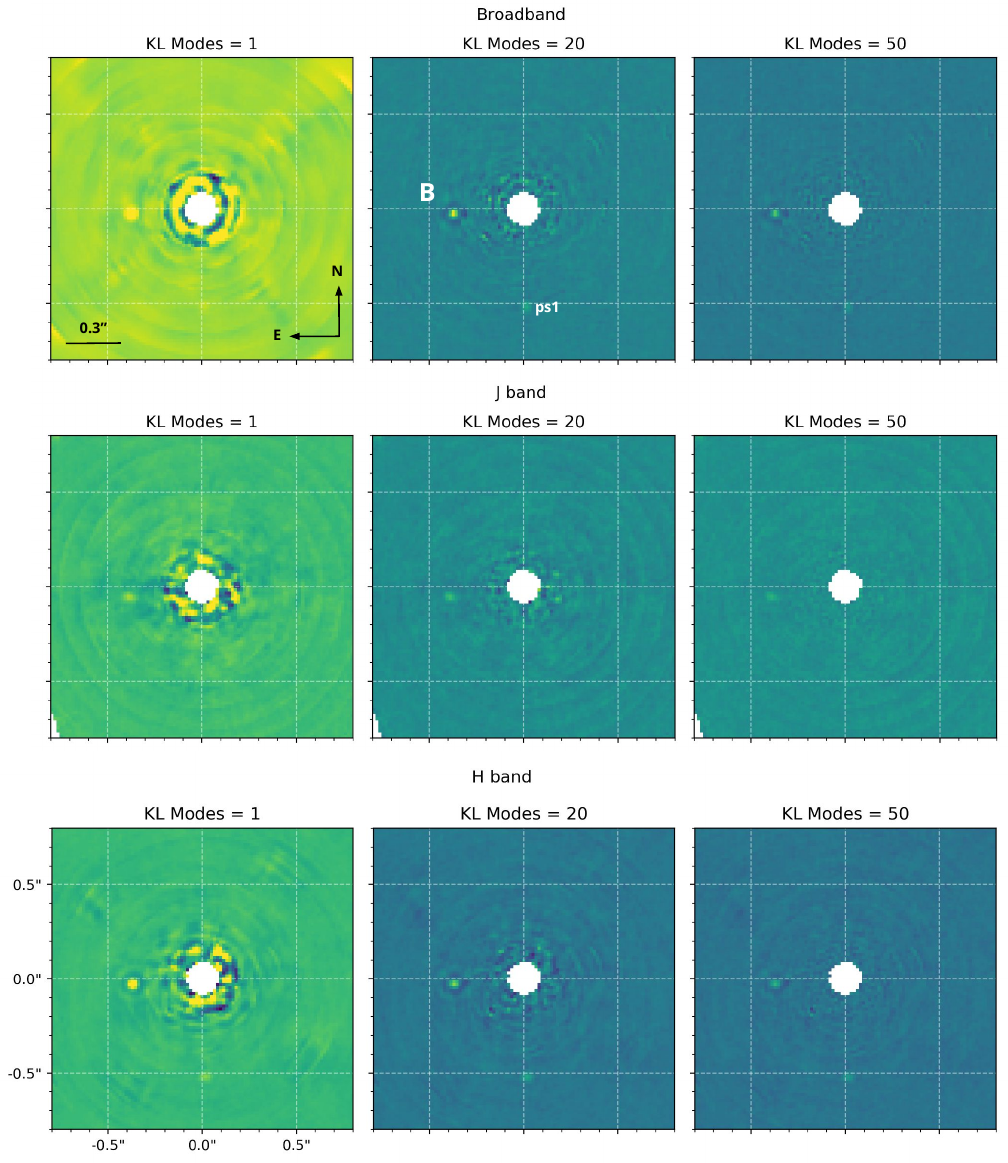}
    \caption{PSF-subtracted images of HIP 93398 B as seen with SCExAO/CHARIS in broadband (top), \textit{J} band (middle), 
    and \textit{H} band (bottom). Each set of images shows the reduction with 1 KL mode, 20 KL modes, and 50 KL modes. The more KL modes included, the higher the risk of oversubtraction. A second, fainter point source (labeled as ps1) is also visible in the 2022 broadband and \textit{H}-band images; this is investigated and found to be a background source in Section \ref{subsec:bg}. J and H band are shown with the same color stretch (maximum value = 20, arbitrary units/detector counts), while broadband uses a different scaling (maximum value = 50).}
    \label{fig:klip-ims}
\end{figure*}

First, \texttt{pyklip-CHARIS} uses the satellite spots to locate the image centroid in preparation for Karhunen-Lo\'{e}ve Image Processing (KLIP) \citep{soummer2012KLIP}. We perform KLIP using both ADI and SDI with the following parameters: 9 annuli, 4 subsections, 1 pixel movement, and 1, 20, and 50 as the maximum number of basis vectors used in noise reconstruction for PSF subtraction. The KLIP PSF-subtracted images are shown in Figure \ref{fig:klip-ims}. 

After the initial KLIP reduction, we perform forward modeling (KLIP-FM, \citet{Pueyo2016ApJ,Wang2016AJ}) to model the companion PSF and fit for its astrometry, testing 1, 20, and 50 KL modes to find a balance between oversubtraction and sufficient noise removal. Using an estimate of the companion's location and a known instrumental PSF, KLIP-FM produces the PSF-subtracted image from data and a forward-modeled companion PSF at the estimated location. We then convert the PSF model flux to the flux of the unocculted star using the satellite spot flux and a known reference-spot-to-star ratio as reported in \citet{currie2020sky}. Pending calibration results from the CHARIS team indicate that this ratio may not be stable over time, and as a result we use the latest value for data taken with the default satellite spot modulation frequency after May 2022 of 1.25$\times 10^{-3} \pm 1.30\times 10^{-4}$ (Kellen Lawson \& Thayne Currie, priv. comm., \citet{currie2020sky}). 

The forward model is then fit using a Markov-Chain Monte Carlo (MCMC; implemented in \texttt{emcee} from \citealt{Foreman-Mackey2013PASP}) to approximate the posteriors of the RA and Dec offsets, a flux scaling factor for the model, the companion separation, and the position angle (PA). For this MCMC implementation, we use uniform priors, 100 walkers with 2000 steps, 400 of which are discarded as burn-in. The maximum likelihood values of the posteriors from the MCMC are adopted as the raw astrometric values, i.e.~before including instrumental calibration uncertainties. For CHARIS, we use uncertainties of 0.05 mas (30\% of a lenslet) on the position of the central star, a lenslet scale of 16.15$\pm$0.10 mas/lenslet \citep{kuzuhara2022direct}, the uncertainty of the aforementioned satellite spot/host star contrast, and a PA offset of $2.\!\!^\circ0 \pm0.\!\!^\circ4$ \citep{Chen2023RASTI}. 

\begin{figure*}
    \centering
    \includegraphics[width=0.75\textwidth]{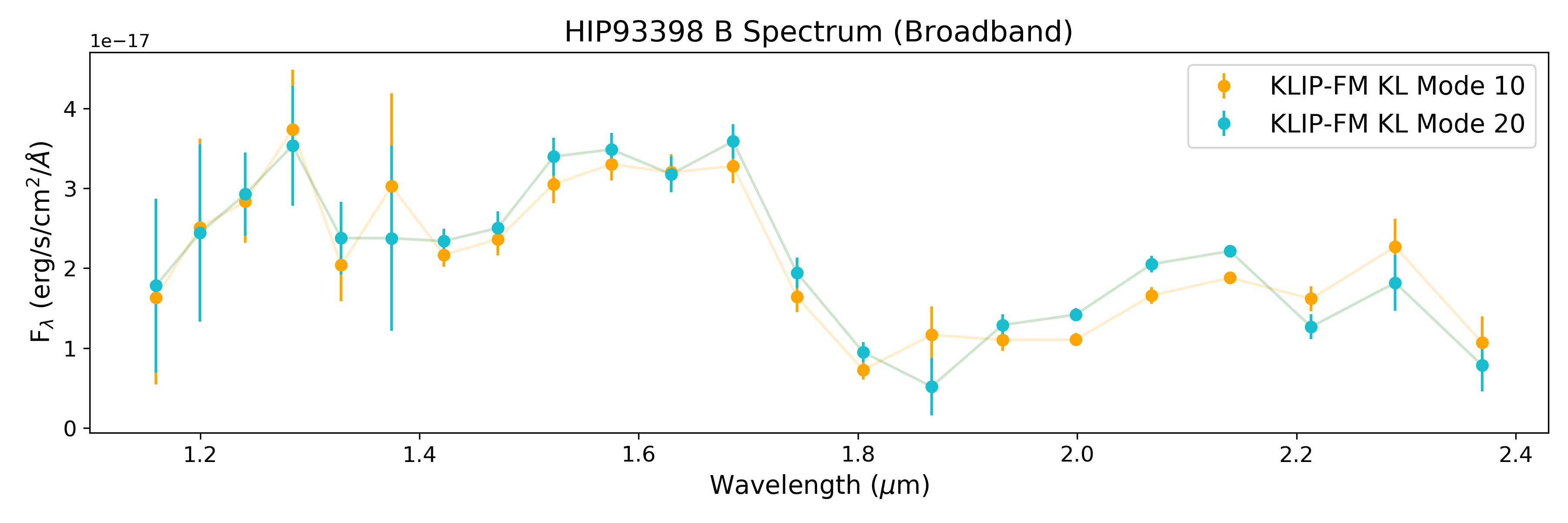}
    \includegraphics[width=0.9\textwidth]{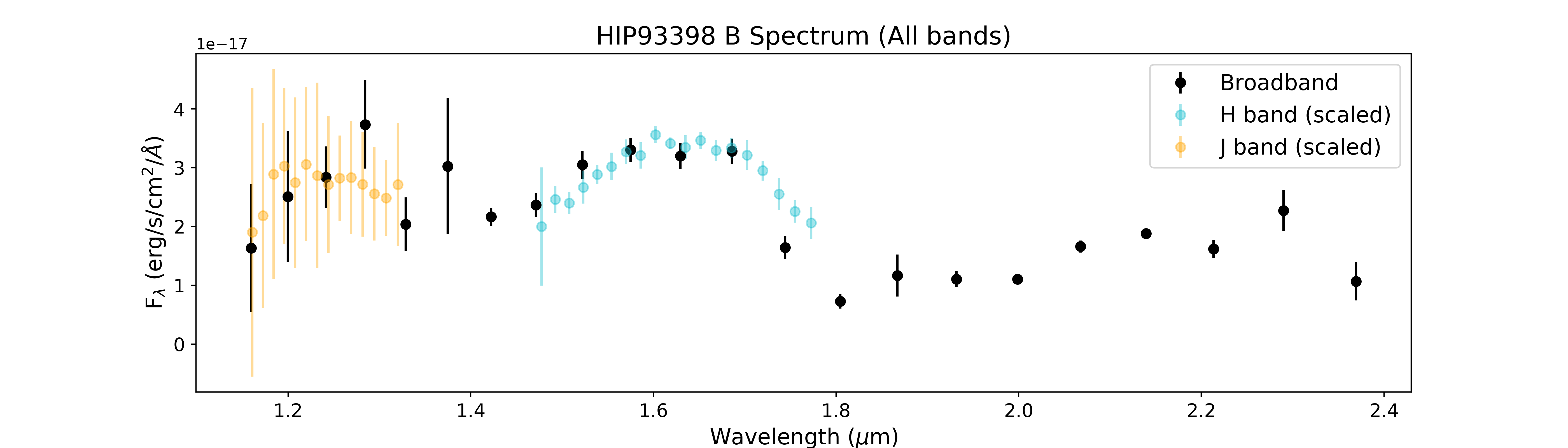}
    \caption{Extracted spectra of HIP 93398 B from the broadband and high-resolution \textit{H}-band data. Methane absorption features are visible at 1.6 and 2.2 $\mu$m. (Top) Two extractions are shown: one using 10 KL modes in the reduction (orange), and one using 20 KL modes (blue). (Bottom) Broadband data with \textit{J}- and \textit{H}-band high resolution superimposed on top, and scaled to near the broadband flux calibration \rev{for illustration purposes}. \rev{It is worth noting that the \textit{J} and \textit{H} band data points are not necessarily directly comparable to the broadband data points, as these are monochromatic flux densities over a finite bandpass and no color correction has been applied.}} 
    \label{fig:spec}
\end{figure*}

KLIP-FM also allows us to extract the companion's spectrum, which, for the broadband data, we then convert from contrast to physical units. This requires the observed magnitude of the host star (listed in Table \ref{tab:obs}), the contrast between the unocculted star and our PSF model, and a stellar model spectrum for the host star, for which we use a Castelli/Kurucz model \citep{Castelli2003IAUS} implemented in \texttt{pysynphot} \citep{STScIDevelopmentTeam2013ascl} with stellar parameters $T_{\rm eff} = 4727$\,K, $[{\rm Fe/H}] = -0.15$, and $\log g  = 4.64$ as referenced in Table \ref{tab:sysprops} above. This calibration can be expressed as
\begin{equation}
    F_{\rm companion} = \frac{F_{\rm companion}}{F_{\rm spot}} \times \frac{F_{\rm spot}}{F_{\rm star}} \times F_{\rm star},
\end{equation}
where $F_{\rm companion}/F_{\rm spot}$ is the raw contrast spectrum previously extracted, $F_{\rm spot}F_{\rm star}$ is the star-to-spot ratio previously used, and $F_{\rm star}$ is the stellar model flux calibrated to the observed magnitude. This results in a calibrated spectrum in flux density units, $F_\lambda$ (erg\,s$^{-1}$\,cm$^{-2}$\,\AA$^{-1}$). Error bars on this spectrum are estimated by injecting and recovering synthetic sources at an annulus of the same separation as the companion, resulting in the spectrum shown in Figure \ref{fig:spec}. We do not report an absolute flux calibration for \textit{J} band, due to the poorer quality of the \textit{J}-band data \rev{resulting from variable seeing during those observations}. Additionally, the \textit{H}-band absolute flux calibration differs slightly from the broadband calibration. We have included this difference in the error on our reported H band magnitude; however, for color measurements, we use \textit{H} and \textit{J} both derived from broadband, as the relative calibration between bands in that observation is more reliable. In Figure \ref{fig:spec}, include both \textit{H} and \textit{J} spectral data with an arbitrary scaling factor to compare their overall shape with that of the calibrated broadband and H band spectra\rev{. Subtle differences in the spectra as shown in Figure \ref{fig:spec} should be interpreted with caution, as no color correction has been applied to account for the different bandpasses between the higher-resolution and broadband modes; additionally, the discrepancy between the \textit{H} band and broadband data is larger than it appears in Figure \ref{fig:spec}, but is included in the error budget on our derived photometry}.

\subsection{Background Object Identification}\label{subsec:bg}

\begin{figure*}
    \centering
    \includegraphics[width=0.75\textwidth]{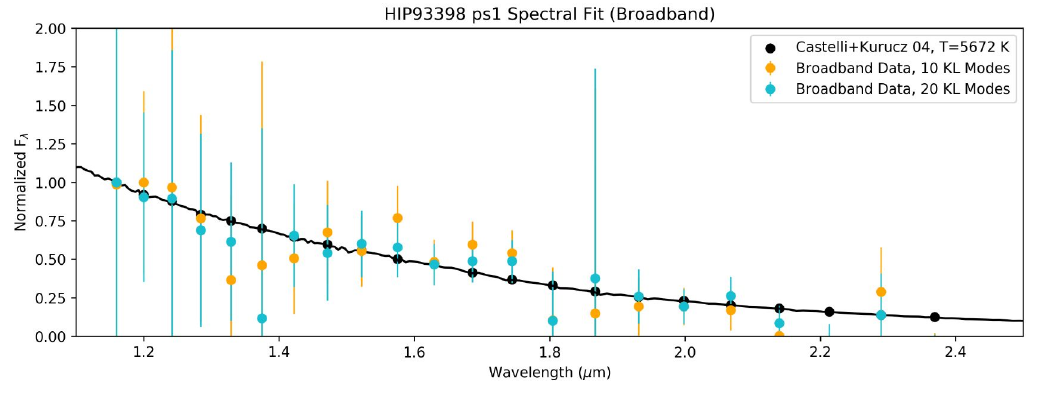}
    \caption{Extracted spectrum of HIP 93398 C from our SCExAO/CHARIS broadband data shown with Castelli-Kurucz 04 model stellar atmosphere fits \citep{castelli2004new}. The best fit model atmospheres indicate a stellar object with a temperature between 3800-5700 K.} 
    \label{fig:ps1}
\end{figure*}

An additional point source with an \textit{H}-band apparent magnitude of $\sim$17 was detected in the 2022 data set at $180^\circ \pm 5.0\!
\!^\circ4$ PA and $515\pm 15.5$ mas separation in the broadband and \textit{H}-band images. We believe this to be a background object, based on its spectrum as shown in Figure \ref{fig:ps1}. We perform a least squares fit to a grid of normalized Castelli-Kurucz model stellar atmospheres \citep{castelli2004new} for the broadband and \textit{H}-band data each separately, as well as together using all available data. The best fit model ($\chi_\nu^2 =0.13$) to the broadband data is a star with $T_{\rm eff} = 5672$\,K, solar metallicity, and $\log g = 4.95$. As a result, we conclude that this is likely a background star with temperature 3800-5700 K. 

Using our broadband data, we determine an observed apparent \textit{K}-band magnitude of 18.6$\pm$1.37. We use this apparent \textit{K}-band magnitude to determine a distance estimate using the surface brightness color relation, as described in \citet{pietrzynski2019distance}. For this calculation, we assume a $V-K$ color (1.53) and radius (0.93$R_\odot$) typical for a star near the best fit temperature \citep{cox2015allen,zombeck2006handbook}, and derive a distance of $\approx$10.4 kpc, a reasonable distance for a background star in the direction of the Galactic center.

Given the lack of a second detection in the 2023 dataset, we are also unable to establish common proper motion for this point source with the host star; however, due to the poor seeing conditions of the 2023 \textit{J}-band epoch, we also cannot use this non-detection to rule out common proper motion.

\section{Orbital Fitting}\label{sec:orbit}

\citet{Li2023MNRAS} indicated a need for future data to further refine the orbital constraints for HIP 93398 B. Our recent epoch of data from 2023 Sept 1 is sufficiently removed from the observations in \citet{Li2023MNRAS} that the object would have shown substantial motion ($\approx$2-3$^\circ$ PA and 0.2'' separation); accordingly, we perform a joint fit using our new epoch of relative astrometry from Subaru/CHARIS, existing Keck/NIRC2 relative astrometry \citep{Li2023MNRAS}, HARPS radial velocities (RVs) \citep{Trifonov2020AandA}, and stellar absolute astrometry from the HGCA \citep{Brandt2018ApJS,Brandt2021ApJS,GaiaCollaboration2020AandA,GaiaCollaboration2022AandA}.
We do not include our 2022 Subaru/CHARIS astrometry in the fit, as it does not provide additional leverage and, although consistent, has larger errors than the existing 2022 Keck/NIRC2 relative astrometry.

\begin{figure*}
    \centering
    \includegraphics[width=0.35\textwidth]{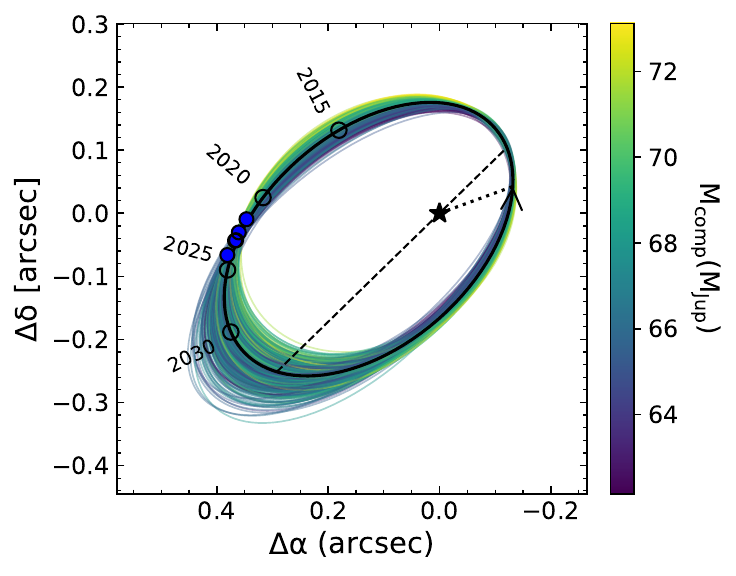}
    \includegraphics[width=0.30\textwidth]{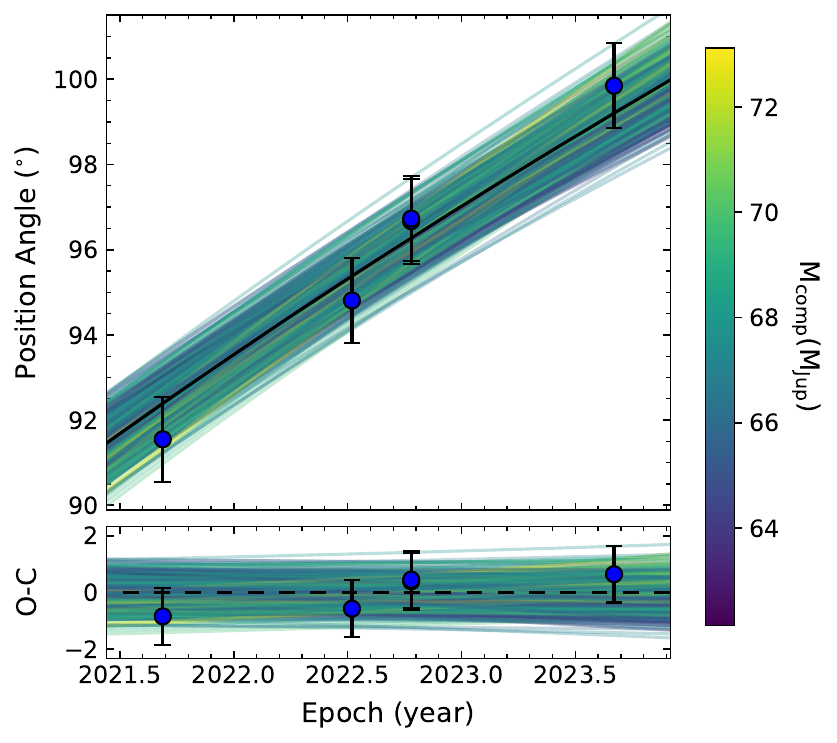}
    \includegraphics[width=0.31\textwidth]{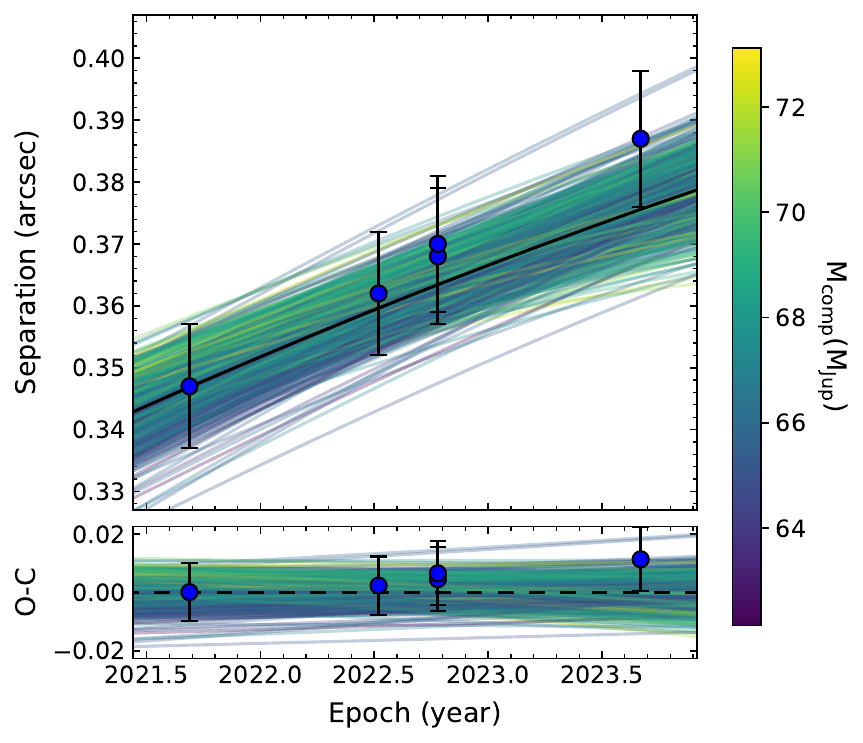}
    \includegraphics[width=0.65\linewidth]{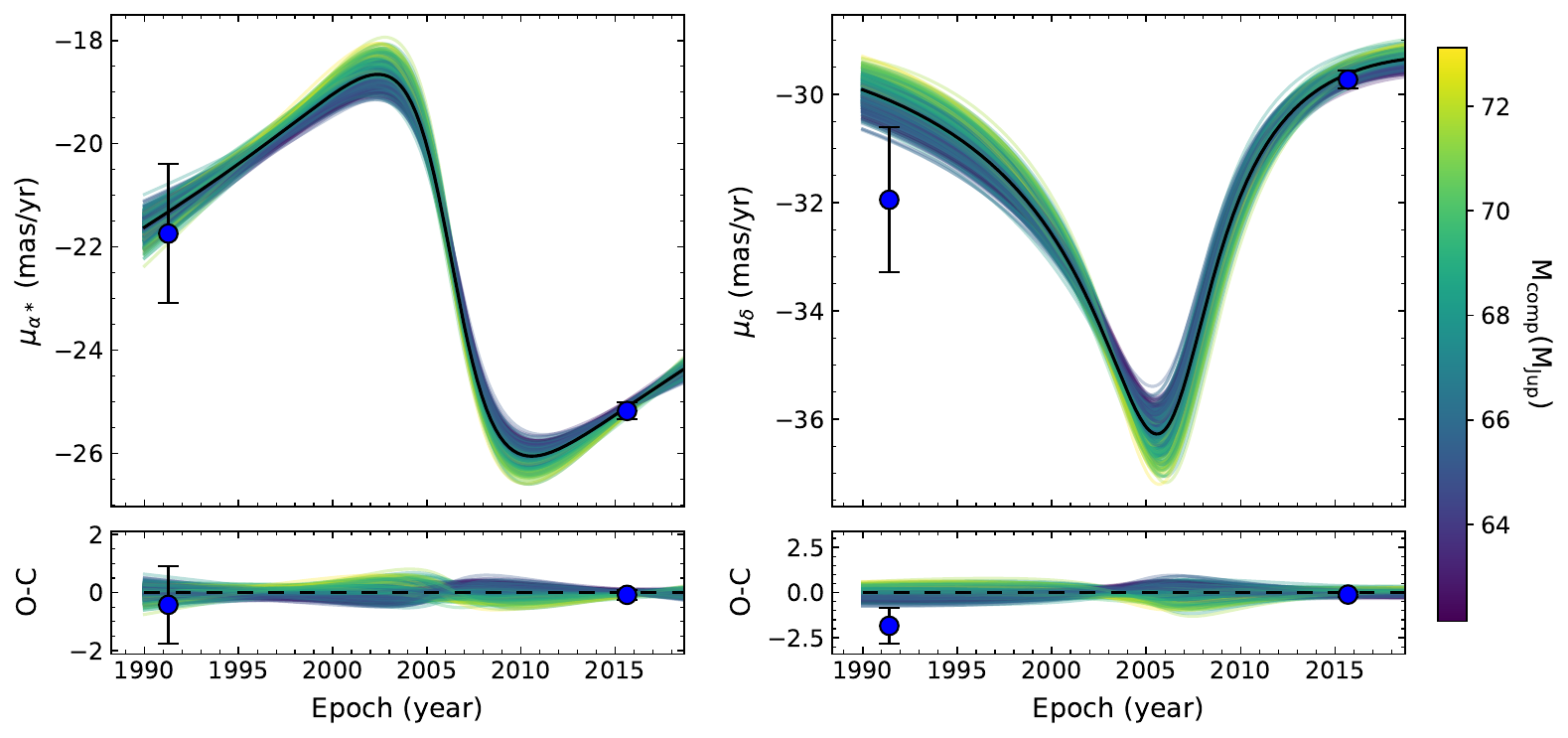}
        \caption{Orbital fits for HIP 93398 B using \texttt{orvara} \citep{brandt2021orvara}. Randomly drawn orbits from the MCMC chain are color-coded by companion mass and plotted against (top) the object's apparent position, measured position angle, and measured separation; and (bottom) absolute astrometry \citep{Brandt2021ApJS}. 
        A companion mass of $\approx$66\,$M_{\rm Jup}$ is favored, in agreement with past work, cementing this object firmly in the brown dwarf regime.
    \label{fig:orvaraorbit}}
\end{figure*}

We fit this combined data using the Bayesian orbit fitting code \texttt{orvara} \citep{brandt2021orvara}, which fits Keplerian orbits using the \texttt{ptemcee} parallel-tempered affine-invariant MCMC algorithm \citep{2016MNRAS.455.1919V,Foreman-Mackey2013PASP} and nine parameters describing the companion's orbit: semi-major axis ($a$), inclination ($i$), longitude of ascending node ($\Omega$), time of periastron passage ($T_0$), eccentricity and argument of periastron ($\sqrt{e}\sin\omega$ and $\sqrt{e}\cos\omega$), companion (secondary) mass ($M_s$), and host star mass ($M_*$). We use log-uniform priors for semi-major axis and companion mass, a geometric prior for inclination, and a Gaussian prior for host star mass based on previously published stellar mass values of $0.72 \pm 0.02$ $M_\odot$. For this MCMC, we use 10 temperatures, 100 walkers, 200,000 steps, and discard 500 of those steps as burn-in. Figure \ref{fig:orvaraorbit} illustrates the orbital fit for HIP 93398 B using \texttt{orvara}, compared with both absolute and relative astrometry. Our orbital fit is in agreement with that of \citet{Li2023MNRAS} within 1$\sigma$.

As a check for consistency, we additionally fit for the orbit of HIP 93398 B using the HGCA module of \texttt{orbitize!} \citep{2020AJ....159...89B}, an alternative open-source orbit fitting code for directly imaged objects. By default, \texttt{orbitize!} uses uniform priors, except for semi-major axis (log uniform), inclination (sine), parallax (Gaussian), companion mass (log uniform), and host star mass (Gaussian). We use\texttt{orbitize!}'s default priors (except on the stellar mass, for which we use the above Gaussian prior), \texttt{ptemcee}, and the above data, except for the HARPS RVs. For clear comparison to the results from \texttt{orvara}, this parallel-tempered MCMC also used 10 temperatures, 100 walkers, 500 burn-in steps, and 200,000 total steps in the chain. Values from \texttt{orbitize!} are in agreement within error bars with both the \citet{Li2023MNRAS} values and our \texttt{orvara} fit results. We report updates to a subset of the parameters fit by \texttt{orvara} and \texttt{orbitize!}--M$_s$, $a$, $e$, and $i$--and compare them to the constraints from \citet{Li2023MNRAS} in Table \ref{tab:orbitfit}. Although this additional epoch of relative astrometry only marginally improves constraints on the system's orbit and companion mass, we confirm that this object is firmly in the brown dwarf regime at $\sim$66 $M_{\rm Jup}$, with a semi-major axis of $\approx$11 au and moderate eccentricity around $\approx$0.5 in line with results for the eccentricity distribution of directly imaged brown dwarfs from \citet{bowler2020population}.

\begin{figure*}
    \centering
    \includegraphics[width=0.8\linewidth]{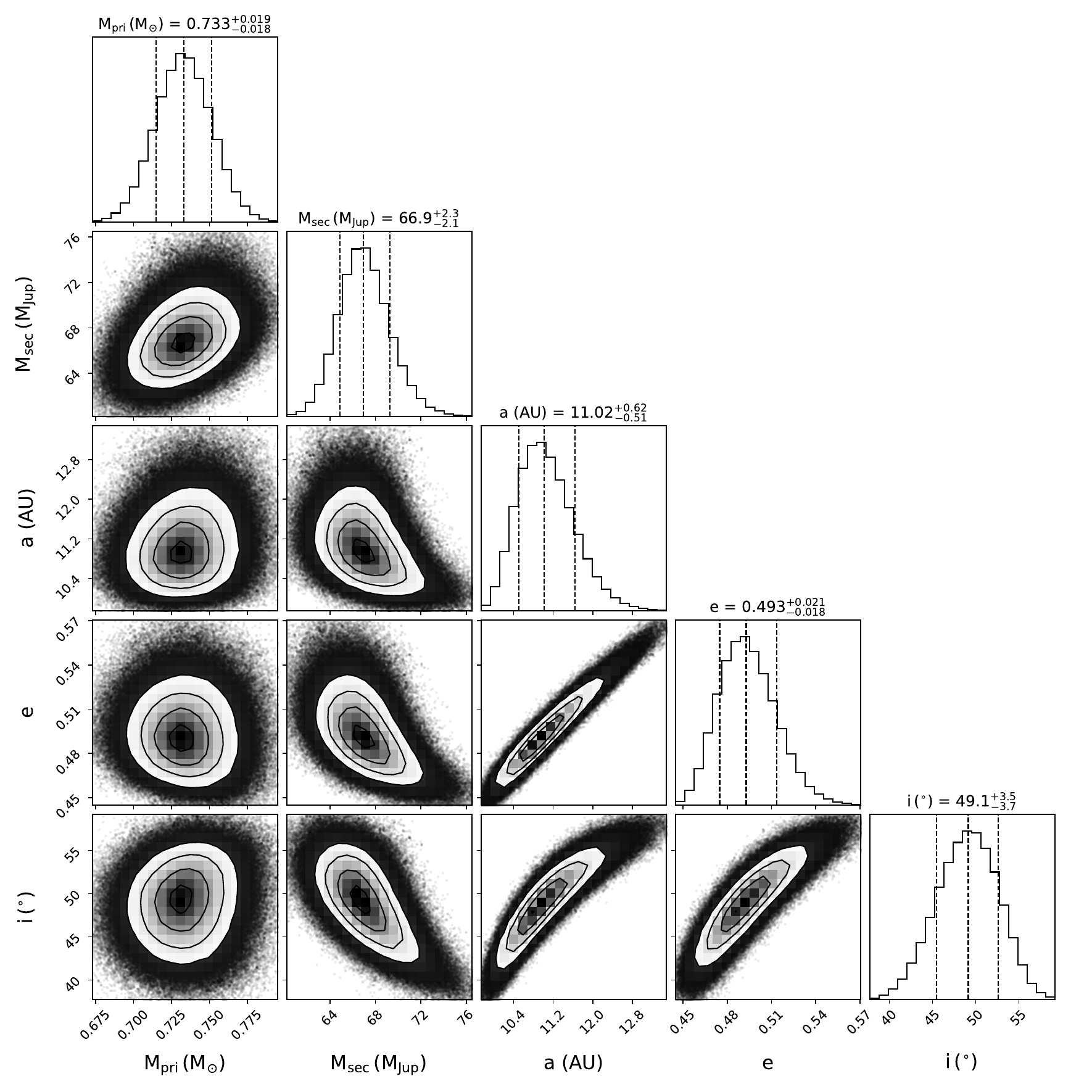}
    \caption{Corner plot for a subset of fit parameters from \texttt{orvara}: stellar mass, companion mass, semimajor axis, eccentricity, and inclination. The posteriors are nearly Gaussian.}
    \label{fig:orvaracorner}
\end{figure*}

\begin{deluxetable*}{lccc}
\tablewidth{0pt}
\tablecaption{Comparison of selected HIP~93398~B parameters from orbit fitting \label{tab:orbitfit}}
\tablehead{
Parameter &
\cite{Li2023MNRAS} &
This Work (orvara) &
This Work (orbitize!)
}
\startdata
Mass ($M_{\rm Jup}$) & $65.9^{+2.0}_{-1.7}$     & $66.9^{+2.3}_{-2.1}$        & $60.3^{+18.8}_{-8.1}$          \\ 
Semimajor Axis $a$ (au)                        & 11.05$^{+0.64}_{-0.56}$  & 11.02$^{+0.62}_{-0.51}$     & 12.17$^{+1.8}_{-1.2}$         \\ 
Eccentricity $e$  & $0.50^{+0.022}_{-0.022}$ & $0.49^{+0.021}_{-0.018}$    & $0.49^{+0.11}_{-0.12}$        \\ 
Inclination $i$ ($^\circ$)                    & 49.8$^{+3.4}_{-3.7}$     & 49.1$^{+3.5}_{-3.7}$        & 56.6$^{+12.9}_{-11.7}$          \enddata
\end{deluxetable*}

\section{Spectral Analysis}\label{sec:spec}

\rev{Using the age of the system and HIP 93398 B's mass as listed in Table \ref{tab:sysprops}, evolutionary models predict the companion to have a temperature of $\sim$1200--1400 K and a luminosity $log(L_{\rm bol}/L_\odot)$ of $\sim$-4.5. This is in tension with the values derived by \citet{Li2023MNRAS}, where} $log(L_{\rm bol}/L_\odot)$ \rev{= -5.26$\pm$0.07 and} $T_{\rm eff}\sim$\rev{900--1000 K, which is fainter and cooler than predicted by evolutionary models. In this section, we extract \textit{J}, \textit{H}, and \textit{K} band photometry to compute a revised bolometric luminosity, and then use our new spectral observations to determine a \revv{revised temperature and} spectral type for the object.}

\subsection{Photometry and Colors}\label{subsec:phot}

\textit{J}-, \textit{H}-, and \textit{K}-band photometry are primarily derived from the calibrated broadband spectrum, due to the more reliable relative calibration between the two channels in that observation, which will be crucial for determining colors. We compute the photometry in each band by integrating the observed flux multiplied by the transmission profile for each filter (\textit{J}$_\textup{MKO}$, \textit{H}$_\textup{MKO}$, and \textit{K}$_\textup{MKO}$, from \citet{rodrigo2020svo}) over its bandpass via the following equation as in \citet{tokunaga2005mauna},
\begin{equation}
    \langle F_{\nu}\rangle = \frac{\int F_\nu(\nu) S(\nu)/\nu \; d\nu}{\int S(\nu)/\nu \; d\nu},
\end{equation}
where $\langle F_{\nu}\rangle$ is the monochromatic flux density (erg/s/cm$^2$/Hz) for the resulting photometry, $F_\nu$ is the observed monochromatic flux density (erg/s/cm$^2$/Hz), $S(\nu)$ is the filter transmission function, and $\nu$ represents the frequencies in each bandpass. Monochromatic flux densities are then converted to Vega magnitudes using the MKO filter zero points. As mentioned earlier in Section \ref{sec:data}, we do not attempt an absolute flux calibration for the \textit{J}-band high-res observations due to data quality, and we incorporate the discrepancy between the \textit{H}-band as measured in broadband and as measured in high-res mode as part of the error on the reported photometry.

\begin{figure*}
    \centering
    \includegraphics[width=0.45\textwidth]{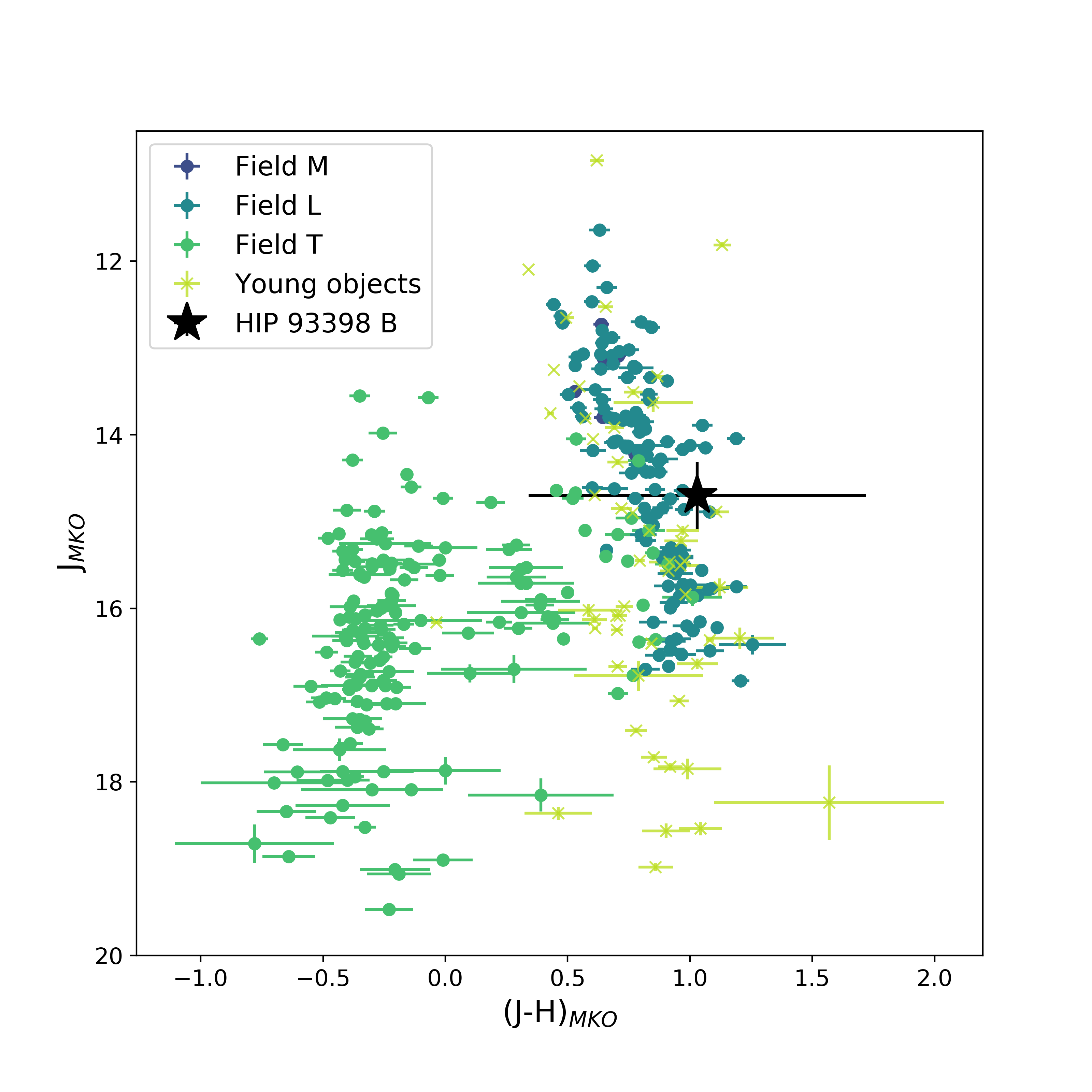}
    \includegraphics[width=0.45\textwidth]{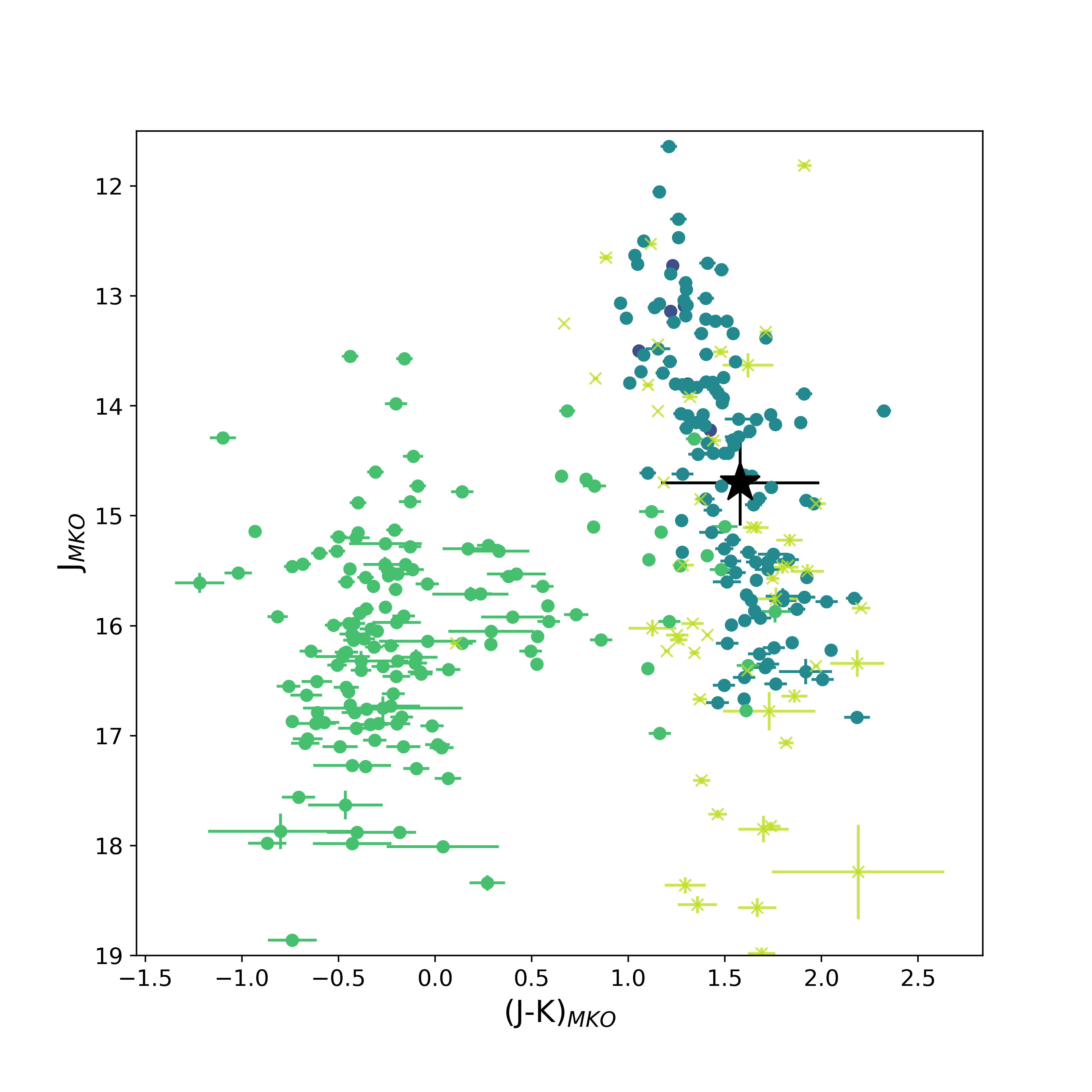}
    \includegraphics[width=0.45\textwidth]{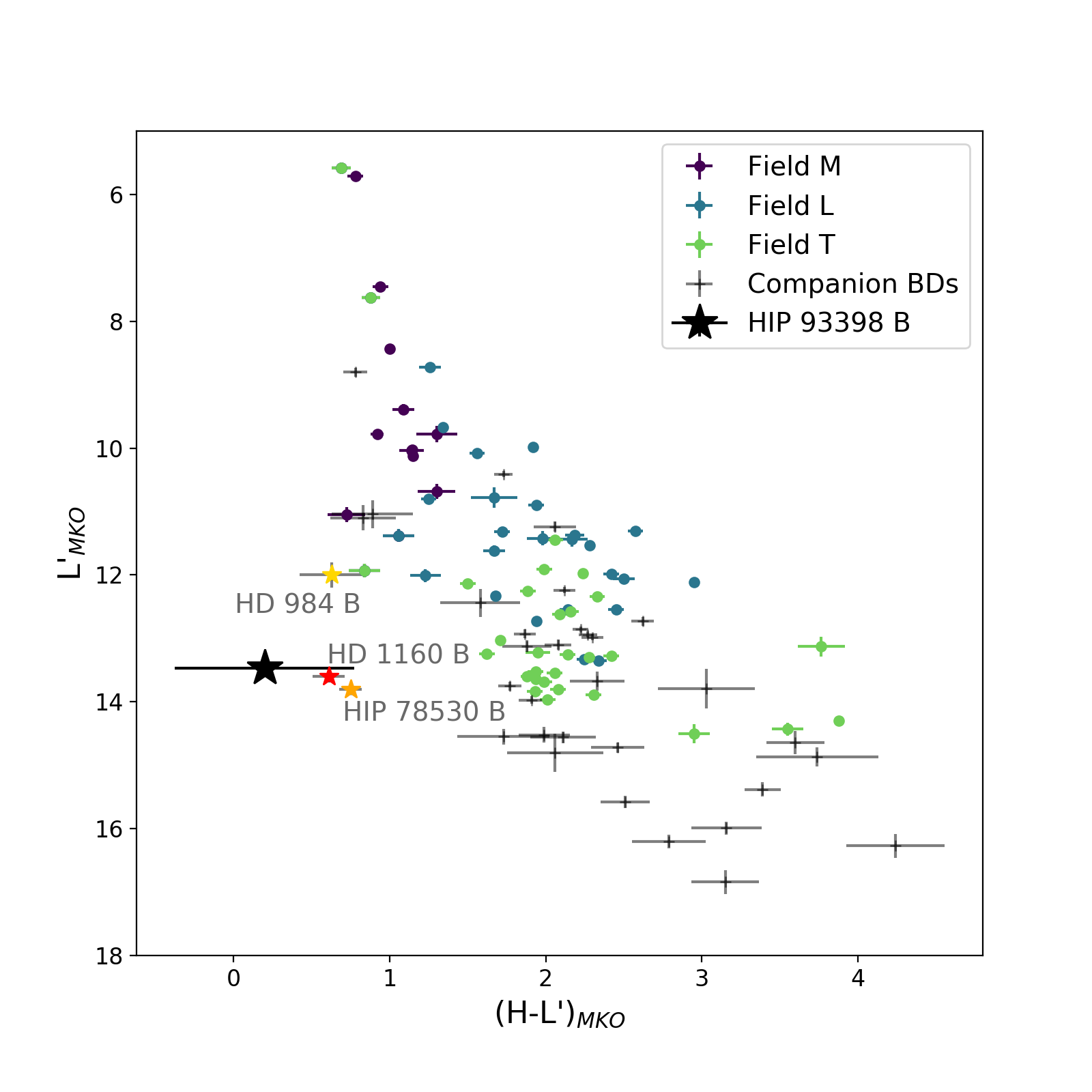}
    \caption{Color magnitude diagrams for $J-H$, $J-K$, and $H-L'$ colors. HIP 93398 B is shown in the context of a number of other field dwarfs and young companion brown dwarfs, as reported in the Ultracool Sheet \citep{best_2024_10573247}. HIP 93398 B is aligned with other L/T transition objects in $J-H$ and $J-K$ color space, while in $H-L'$ it is an outlier, somewhat nearby a few other companion BDs, namely HD 984 B (yellow, \citet{2015MNRAS.453.2378M,2016ApJ...833...96L}), HD 1160 B (red, \citet{2016AandA...587A..56M,2017ApJ...834..162G,2012ApJ...750...53N}), and HIP 78530 B (orange, \citet{2013ApJ...767...31B,2015ApJ...802...61L,2005AandA...430..137K,2011ApJ...730...42L}). The strange $H-L'$ color may be explained by systematic errors in the reported $L'$ magnitude from \citet{Li2023MNRAS}; however, a full re-analysis of those data is beyond the scope of this paper.}
    \label{fig:cmd} 
\end{figure*}

Then, to place HIP 93398 B in context with other substellar objects, we use our newly derived \textit{J}, \textit{H}, and \textit{K} magnitudes, along with the previously published $L'$ magnitude from \citep{Li2023MNRAS}, to compute its $J-H$, $J-K$ and $H-L'$ colors. We determine that $J-H$ = 1.03 $\pm$ 0.69, $J-K$ = 1.58$\pm$0.41 and $H-L' = 0.20 \pm 0.57$ for HIP 93398 B. In Figure \ref{fig:cmd}, HIP 93398 B is shown alongside a sample of field and companion M-, L-, and T-dwarfs from the Ultracool Sheet \citep{best_2024_10573247}. In $J-H$ color space, HIP 93398 B is aligned with other L/T transition objects, while in $H-L'$ it is somewhat of an outlier, along with a few other young companions, namely HD 984 B \citet{2015MNRAS.453.2378M,2016ApJ...833...96L}, HD 1160 B  \citep{2016AandA...587A..56M,2017ApJ...834..162G,2012ApJ...750...53N}, and HIP 78530 B \citep{2013ApJ...767...31B,2015ApJ...802...61L,2005AandA...430..137K,2011ApJ...730...42L}. These three comparison objects have all been identified as much younger ($<$120 Myr) late M-dwarfs, with masses ranging from $\sim$25-125 $M_{\rm Jup}$. Although there is low precision on HIP 93398 B's age measurement, even the lowest bounds indicate this object is at least 500 Myr old. The strange $H-L'$ color alternatively may be explained by systematic errors in the reported $L'$ magnitude from \citet{Li2023MNRAS}; however, a full re-analysis of those data is beyond the scope of this paper.

\rev{Additionally, we derive a bolometric luminosity for HIP 93398 B via the methods described in \citet{kirkpatrick2021field} and \citet{sanghi2023hawaii}. We first use the \citet{kirkpatrick2021field} \rev{polynomial relationship between the object's absolute \textit{H} band magnitude $M_H$ and its effective temperature $T_{\rm eff}$}, resulting in a temperature of 1280$^{+156}_{-138}$ K for HIP 93398 B. Assuming a radius of \revv{R$\sim$0.86$R_{\rm Jup}$ from evolutionary models and using the Stefan-Boltzmann law, this results in a luminosity $log(L_{\rm bol}/L_\odot)$ of -4.78$^{+0.21}_{-0.24}$ dex.} Using the polynomial relationships between bolometric luminosity and \textit{J}, \textit{H}, and \textit{K} magnitudes from \citet{sanghi2023hawaii}, we obtain $log(L_{\rm bol}/L_\odot)$ = -4.65$\pm$ 0.15, -4.64$\pm$0.23, and -4.63$\pm$0.08 respectively. The value we report in Table \ref{tab:sysprops} is that of the \textit{H} band magnitude derived bolometric luminosity from the \citet{sanghi2023hawaii}, as that relationship is described as the most reliable in that work. \revt{It is also worth noting that the luminosity determined using radius and the Kirkpatrick relationship for M$_H$ and $T_{eff}$ agrees to better than one sigma with the luminosity determined using the Sanghi relationships for M$_J$, M$_H$, M$_K$.}}

\subsection{Spectral Standards and Empirical Spectra}\label{sec:analy}
We first compare HIP 93398 B's CHARIS spectra (reduced with 10 KL modes) to spectral standards from the SpeX Prism Library Analysis Toolkit (\texttt{SPLAT}) \citep{Burgasser2017ASInC}. Using \texttt{SPLAT}'s classifyByStandard routine, \revt{we determine that the broadband spectra and the \textit{H}-band spectra are both best matched with L9.0 standard DENIS-P J0255-4700 (\citet{burgasser2006unified}). The mean spectral types calculated by \texttt{SPLAT} are L5.0$\pm$4.5 and L9.0$\pm$0.6 for broadband and \textit{H}-band respectively. The higher uncertainty on the broadband classification in this routine is likely due to the larger errors present in \textit{J} band in the broadband data set; we adopt the \textit{H}-band-derived spectral type of L9.0, which is also consistent with the broadband-derived spectral type within one sigma.}  

\subsection{Comparison to Substellar Atmosphere Model Spectra and Evolutionary Models}

We then fit HIP 93398 B's broadband \textit{J}/\textit{H}/\textit{K} and high-resolution \textit{H}-band CHARIS spectra independently to substellar atmospheres from the Sonora Diamondback \citep{morley2024sonora} model grids. For this portion of the analysis, we omit the high-resolution \textit{J}-band data due to its larger error bars from poorer seeing conditions. 
Sonora Diamondback models are the recently released successor to the well-established Sonora Bobcat models, and include the effects of clouds on substellar spectra (unlike Bobcat, which only included cloud-free models). These models cover effective temperatures from 900 to 2400 K, surface gravities $\log g = 3.5--5.5$, metallicities $[{\rm M/H}] = -0.5$, 0.0, and +0.5, plus cloud sedimentation efficiency $f_{\rm sed}$ \citep{ackerman2001precipitating} values from 1 to 8. Both models are appropriate for a large range of masses of substellar objects. 

\begin{figure*}
    \centering
    \includegraphics[width=\linewidth]{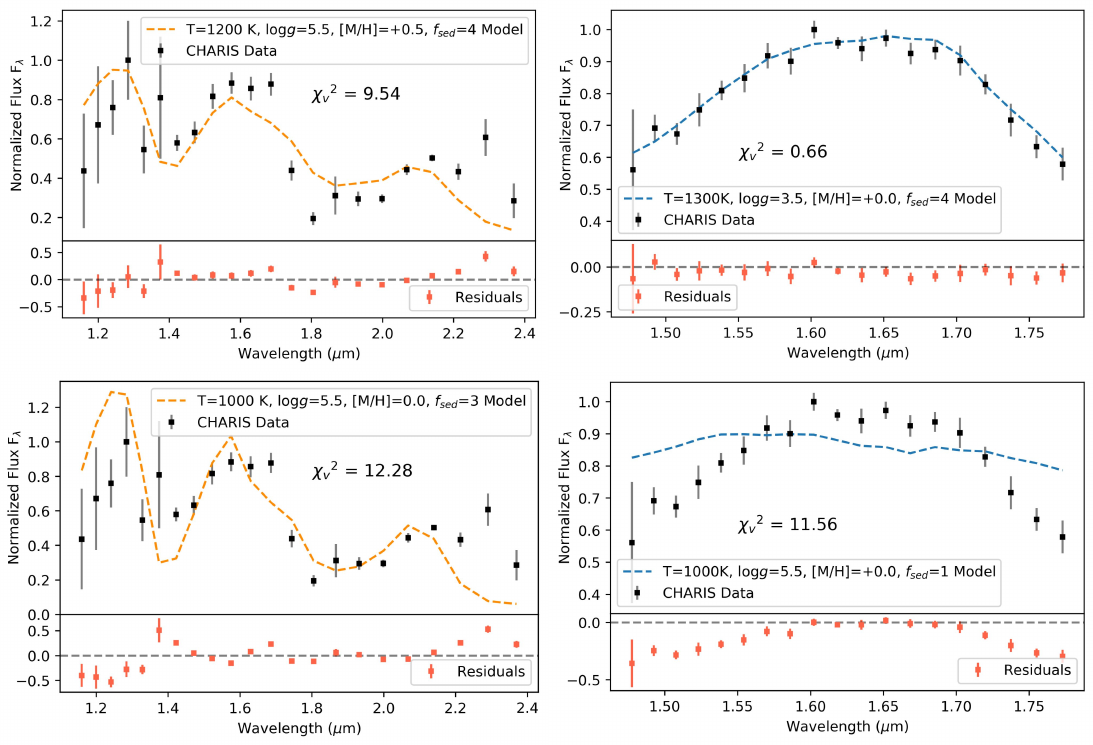}
    \caption{\revt{CHARIS spectra of HIP 93398 B (black) compared to the best fit substellar atmosphere spectral models from Sonora Diamondback \citep{morley2024sonora}. (Top left) The broadband best fit model spectrum with all four free parameters ($\chi_\nu^2$=9.54) is shown in orange, and residuals are shown in red below the spectra. (Bottom left) Here we see the best fit broadband spectrum if we assume solar metallicity and log\textit{g}=5.5 ($\chi_\nu^2$=12.28). (Top right) The \textit{H}-band best fit model spectrum with all four free parameters ($\chi_\nu^2$=0.66) is shown in blue with its residuals. (Bottom right) Here we see the best fit \textit{H}-band spectrum if we assume solar metallicity and log\textit{g}=5.5 ($\chi_\nu^2$=11.56).  The \textit{H}-band fit assuming solar metallicity and log\textit{g}=5.5 is obviously poor; solar metallicity and higher log\textit{g} are not preferred models for the \textit{H}-band data as shown in Figure \ref{spec-model-grid-diamondback}. However, as the \textit{H}-band data only probes a small part of wavelength space and therefore has less leverage to discriminate between models, we adopt the broadband-data derived values as described in text.}}
    \label{spec-model-fits}
\end{figure*}

For comparison to our spectra, we used a Gaussian filter to convolve and resample the model spectra to the CHARIS resolutions for broadband ($R\sim 18.4$) and \textit{H}-band ($R \sim 65$) observations\rev{; the same wavelength region and sampling is used for each model spectrum.} For each spectrum in the model grid, we determine an optimal scaling factor to the data via $\chi^2$ minimization, and then compare \rev{these independent} reduced chi squared $\chi^2_\nu$ values across the grid to find the best fit model. The best fit model for \textit{H}-band has $T_{\rm eff}=1300$ K, $\log g=3.5$, solar metallicity, and cloud parameter $f_{\rm sed}=4$ \revt{when all four parameters are allowed to vary.} The best fit model for the broadband data has $T_{\rm eff}=1200$ K, $\log g=5.0$, metallicity $[{\rm M/H}]=+0.5$, and cloud parameter $f_{\rm sed}=4$ \revt{when all parameters are allowed to vary.} These best fit spectra are shown compared to our data in Figure \ref{spec-model-fits}. 

\begin{figure}
    \centering
    \includegraphics[width=\linewidth]{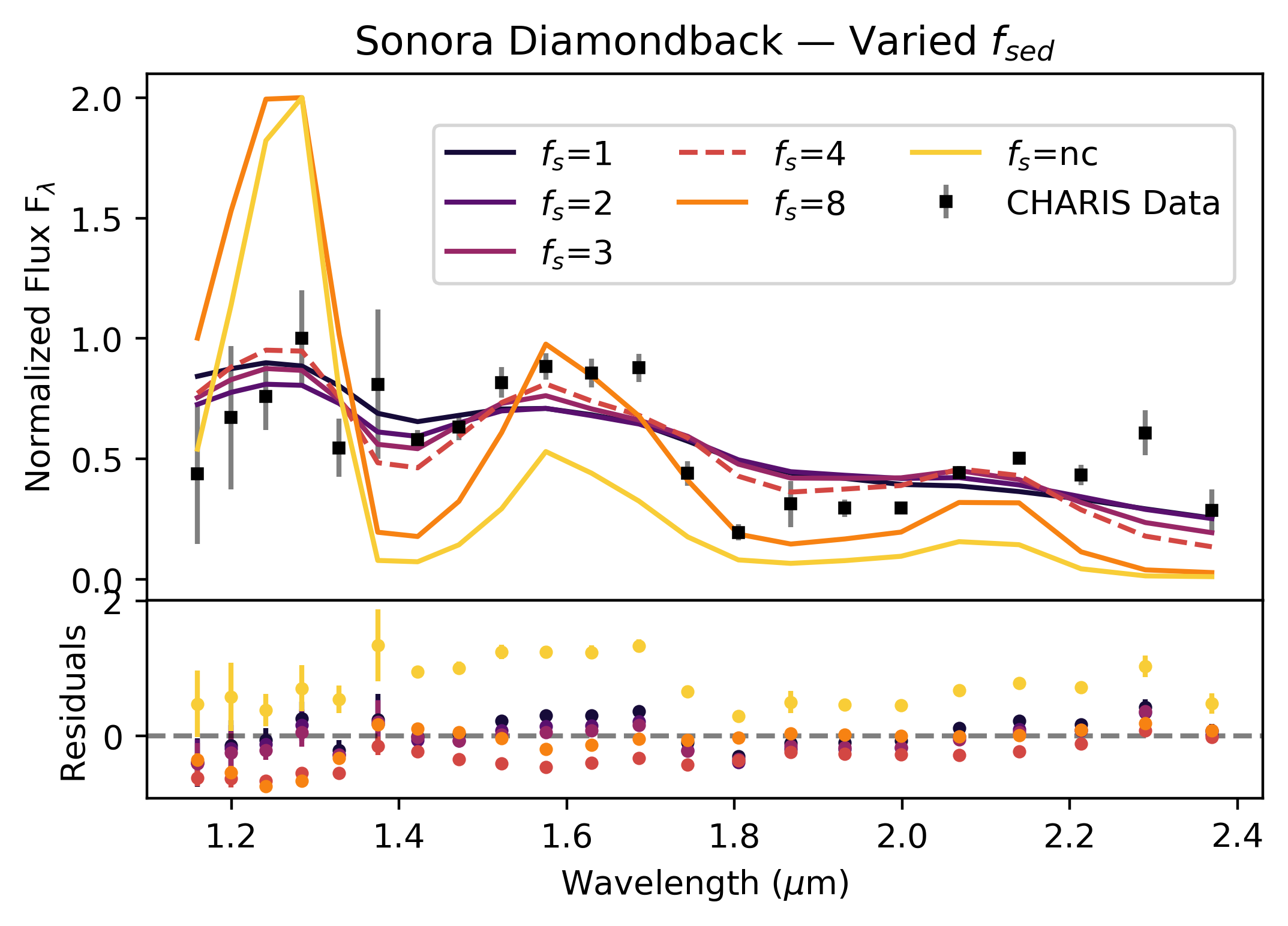}
    \includegraphics[width=\linewidth]{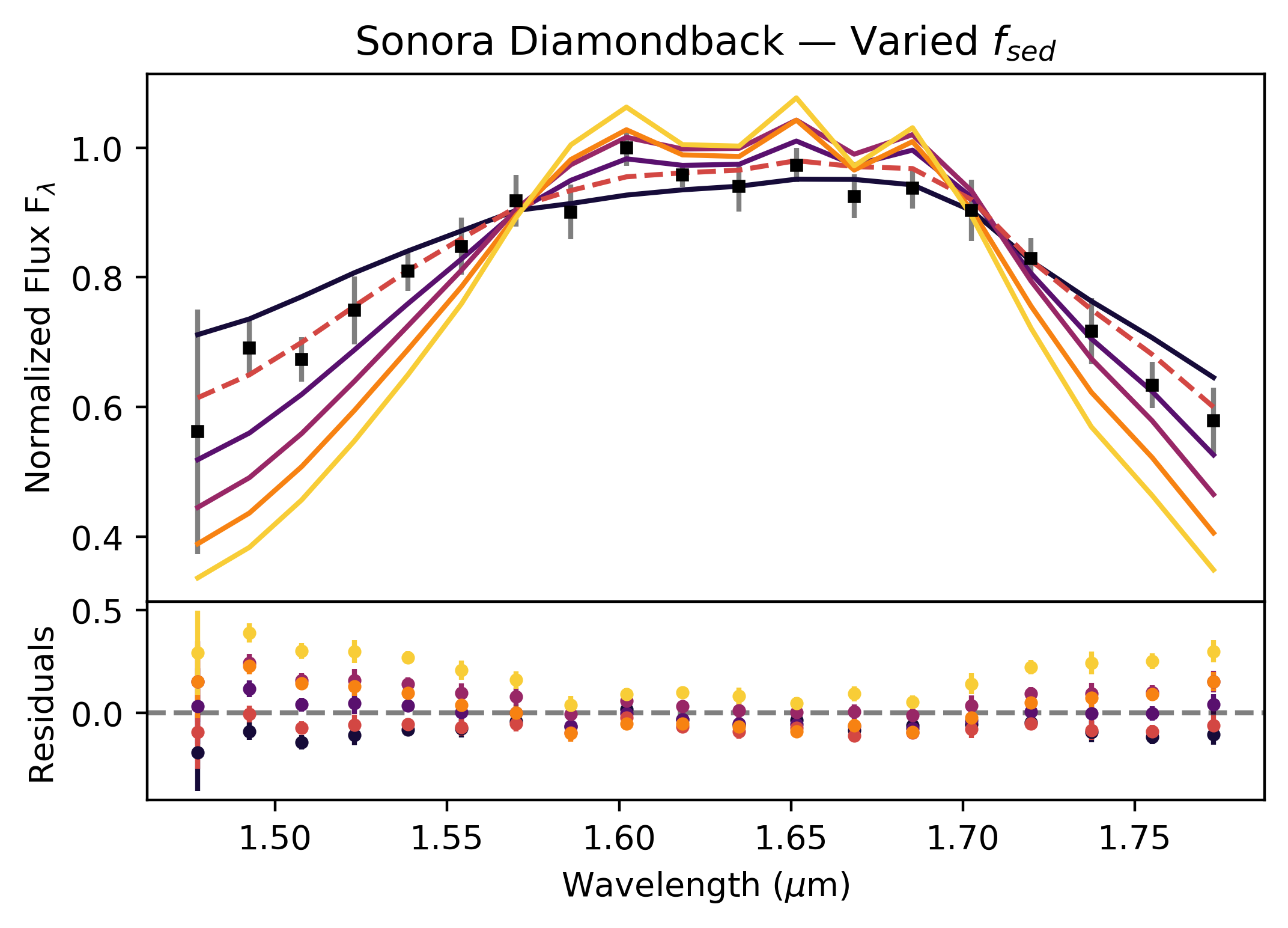}
    \caption{CHARIS spectra of HIP 93398 B (black) compared to the best fit substellar atmosphere spectral models from Sonora Diamondback \citep{morley2024sonora} with varying values of $f_{\rm sed}$ (a parameterization of cloud thickness) at constant temperature, metallicity, and $\log g$ (1200 and 1300 K, +0.5 and 0.0, and 5.0 and 3.5 dex, the best fit values for the broadband and \textit{H} band spectra respectively), where the best fit value is indicated with a dashed line. This object is best fit with models that include clouds with low-to-mid sedimentation efficiency, indicating the \rev{possible} presence of moderate clouds; \rev{although the degree of cloudiness isn't well constrained, clouds are necessary to reproduce the flatness of our observed spectrum.}}
    \label{spec-multifsed}
\end{figure}

\rev{However, it is unlikely for a companion to have a substantially different metallicity from its primary (in this case, near solar metallicity), and for an object of this mass and \revv{R$\sim$0.86 R$_{\rm Jup}$ (as predicted by evolutionary models), we would expect a surface gravity around log$g$=5.5.} \revt{If we assume solar metallicity and log$g$=5.5, the best fit broadband spectrum then has $T_{\rm eff}=1000$ K and $f_{\rm sed}=3$. For those same assumptions, the best fit \textit{H}-band spectrum then has $T_{\rm eff}=1000$ K and $f_{\rm sed}=1$.} \revt{These are also shown compared to the data in Figure \ref{spec-model-fits}.} Additionally, spectra from atmospheres with the same temperature, log$g$, and metallicity but different $f_{\rm sed}$ values are shown in Figure \ref{spec-multifsed}, highlighting the importance of clouds in reproducing the relatively flat spectrum observed for this object.} 

\begin{figure*}
    \centering
    \includegraphics[width=\linewidth]{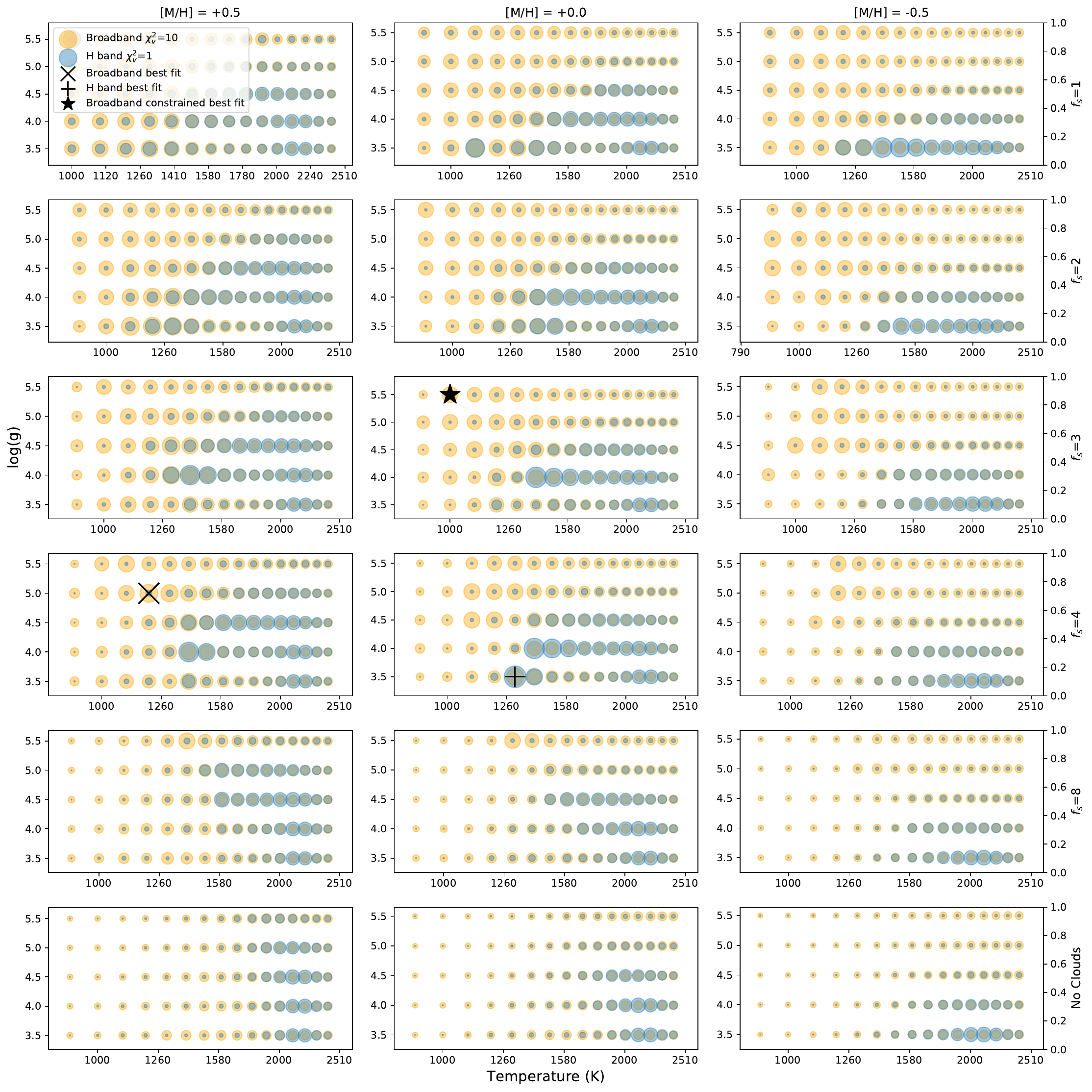}
    \caption{Fit metrics across the Sonora Diamondback model grid for both \textit{H} band (blue) and broadband (orange) CHARIS data for the three available metallicities: +0.5 (left), +0.0 (middle), and $-0.5$ (right). Best fit models are shown in black crosses, \revv{with a black star for the ``constrained'' best fit value, where we assumed solar metallicity and log$g$ = 5.5. A} larger marker size corresponds to a smaller $\chi^2$ value \revt{(i.e.~a better fit, where the size of $\chi^2_\nu = 1$ for \textit{H}-band and the size of $\chi^2_\nu = 10$ for broadband are shown in the legend).} Models without clouds--similar to those from the previous generation of models, e.g. Sonora Bobcat--indicate a higher temperature $>$2000\,K; however, cloudy models and metallicity $\geq$0.0 are significantly favored overall and, for the broadband data, indicate a lower temperature around \revv{1000}--1600\,K, in line with expectations for an L/T transition object and in agreement with evolutionary models, as shown in Figure \ref{evol}.}
    \label{spec-model-grid-diamondback}
\end{figure*}

Maps of $\chi^2$ across the model grid points are also shown in Figure \ref{spec-model-grid-diamondback}, and they illustrate the favored parameter space for each data set; $[{\rm M/H}]<0$ and models with no clouds or thin clouds ($f_{\rm sed}=8$) are clearly disfavored, while a fairly broad range of temperatures are preferred, especially for the \textit{H}-band data. We calculate constraints on temperature from this information by \revv{marginalizing over sedimentation efficiency and surface gravity, assuming solar metallicity, resulting in a temperature of $1200^{+140}_{-119}$ K derived from the broadband data and $2100^{+172}_{-715}$ K from the H band data.} These values agree at the 2$\sigma$ level, and the marginalized posterior distributions show a strongly peaked Gaussian for the broadband fit, but a much broader \revt{skewed and double-peaked distribution, allowing for significantly higher temperatures, for the H band fit, which is visually evident in Figure \ref{spec-model-fits}.} \revt{These values are also in agreement with the effective temperature derived in Section \ref{subsec:phot} from the M$_H$-T$_{\rm eff}$ relatonship from \citet{kirkpatrick2021field}, 1280$^{+156}_{-138}$ K.} Additionally, the \textit{H}-band high-resolution data only probes a small part of wavelength space, and therefore has less leverage to discriminate between models. As a result, we adopt the broadband-data derived temperature and $f_{\rm sed}$ values ($1200^{+140}_{-119}$, $f_{\rm sed}\approx3-4$) for the remaining discussion.

\begin{figure*}
    \centering
    \includegraphics[width=0.65\textwidth]{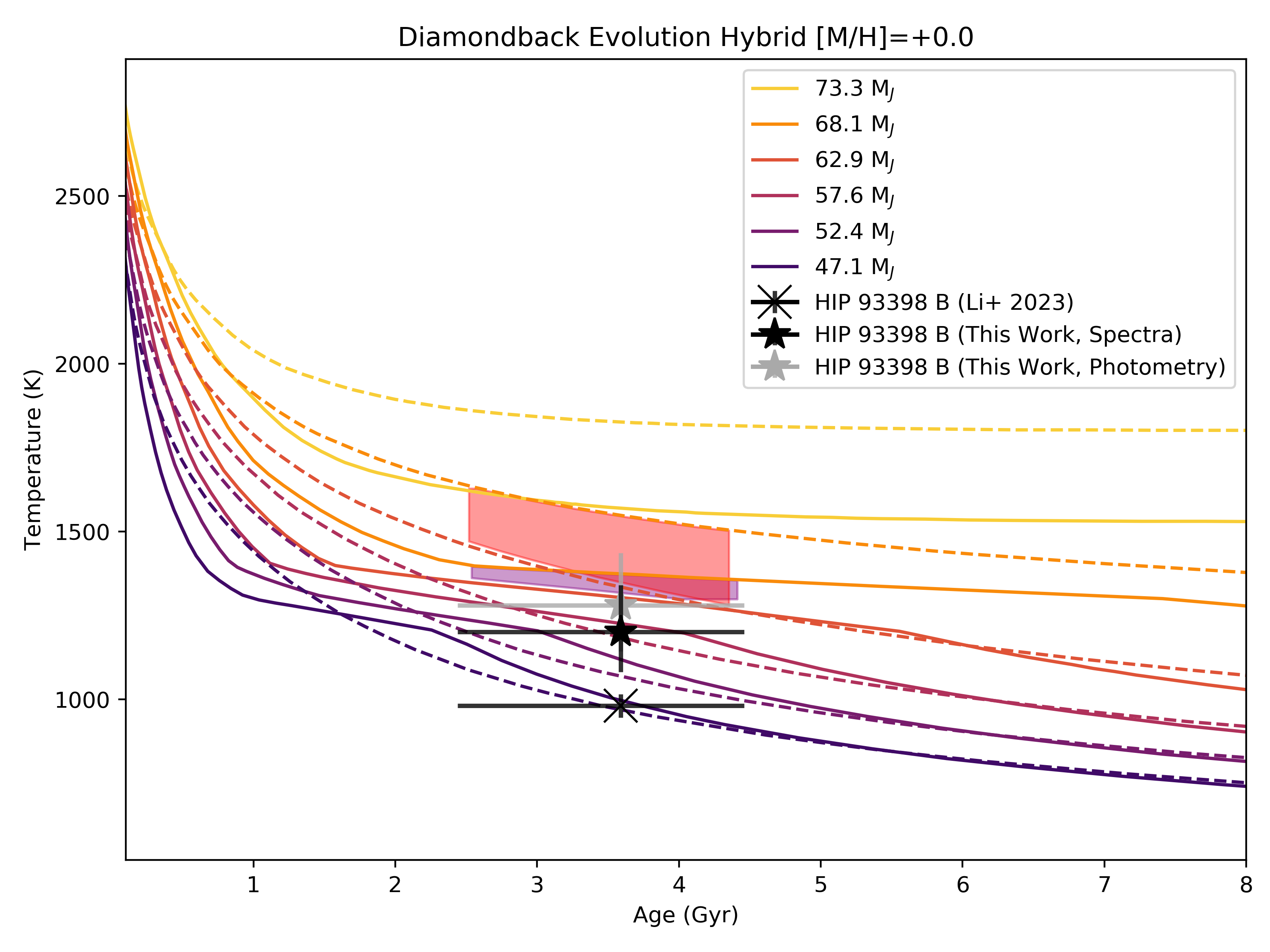}
    \caption{Age and temperature of HIP 93398 B \rev{derived from Diamondback model spectra fitting to \revt{broadband data (1200$^{+140}_{-119}$ K) and from photometric polynomial relationships (1280$^{+156}_{-138}$ K),} compared to Diamondback evolutionary models for solar metallicity. Hybrid models including clouds are shown in solid lines, while cloudless models are indicated with dashed lines.} \revt{Regions consistent with both the independent age and mass constraints are shaded in red for cloudless models, and purple for cloudy models. Our new temperature constraints are consistent with model estimates for an object of this mass and age, while the previous colder temperature estimate is not. Additionally, our spectrum-derived temperature estimate is slightly inconsistent with cloudless models' temperature predictions; however, this discrepancy is quite small.}} 
    \label{evol}
\end{figure*}

\begin{figure*}
    \centering
    \includegraphics[width=0.65\textwidth]{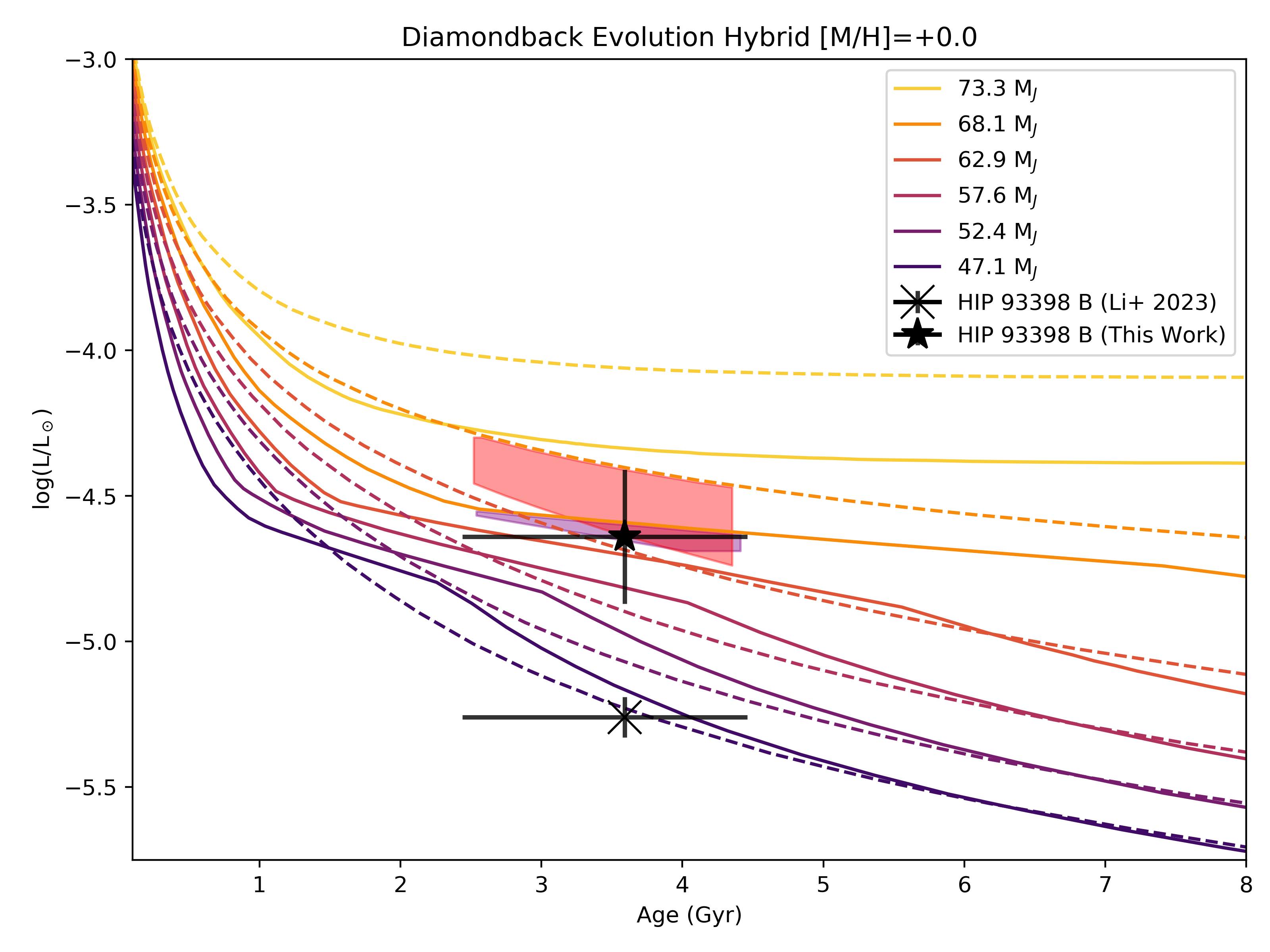}
    \caption{\revt{The same as figure 10, using luminosity instead of temperature for the y-axis. Our luminosity estimate is based on photometry as derived in Section \ref{subsec:phot}. As also shown in Figure 10, the revised luminosity presented herein is consistent with evolutionary models for both cloudy and cloudless atmospheres, while the earlier Li et al. results are not.}} 
    \label{evol2}
\end{figure*}

Additionally, we compare the spectrum-derived temperature of HIP 93398 B and its independently-derived age to evolutionary models for both cloud-free and ``hybrid'' (i.e.~including clouds) models from Sonora Diamondback. As described in \citep{morley2024sonora}, the inclusion of clouds can change the predicted temperature for an object by $\sim$100--200 K and, therefore, has a significant effect on the predicted evolutionary track for a given mass of object. \rev{In Figures \ref{evol} and \ref{evol2}, we place HIP 93398 B in the context of evolution models for various substellar object masses \rev{assuming solar metallicity}. HIP 93398's temperature, spanning $\sim$1000--1500 K when considering estimates from both spectra and photometry, as well as its revised luminosity, are consistent with both cloudy and cloud-free evolutionary tracks for its mass as shown in Figures \ref{evol} and \ref{evol2}. If the temperature is lower as suggested by spectral fits alone, there would be a notable disagreement with cloud-free evolutionary tracks.} The agreement with evolutionary models is also not particularly sensitive to age if the object were to be older, as was estimated by some methods in Section \ref{tab:sysprops}.

\section{Discussion and Conclusions} \label{sec:disc}

\citet{Li2023MNRAS} suggested HIP 93398 B to \revv{be substantially cooler ($<$1000 L)}, a likely T dwarf, based on inference from other analogous substellar objects with similar masses and ages. However, the additional information provided herein from the object's spectra suggest that this is not the case, as HIP 93398 B \revv{more likely has a temperature of \revt{1200$^{+140}_{-119}$ K} and could be classified as a late L dwarf.} Its \revv{revised temperature and luminosity, as well as its location at the L/T transition,} are supported by multiple lines of evidence: \rev{spectral typing via standards}, temperatures derived from substellar atmosphere models and photometry, $J-H$ and $J-K$ colors, and position on color-magnitude diagrams. HIP 93398 B's estimated bolometric luminosity $log(L_{\rm bol}/L_\odot)$ = \rev{-4.64$\pm$0.23} is significantly brighter than the estimate from \citet{Li2023MNRAS}'s $L'$ photometry of -5.26$\pm$0.07. 
Perhaps most notably, modeling fits to this spectrum clearly indicate that \rev{some amount of clouds} are likely present in HIP 93398 B's atmosphere, although further modeling work would be needed to determine the cloud composition \rev{and characteristics}. 


Additionally, the \revv{revised temperature and luminosity for HIP 93398 B} resolve the previously proposed tension between evolutionary models and dynamical masses. \citet{Li2023MNRAS} determined that their T6 classification and dynamical mass measurement justified the addition of HIP 93398 B to a growing list of brown dwarfs that may challenge existing evolutionary models, as they are overmassive and underluminous compared to model predictions. However, our \revv{work} leads to a substantially higher model-dependent $T_{\rm eff}$ and also a higher luminosity, resolving this issue. Using the newest generation of substellar atmosphere/evolution models (Sonora Diamondback; \citet{morley2024sonora}) \rev{--- which include clouds, an important feature} for objects at the L/T transition and a key factor in brown dwarf atmospheres \citep{marley2015cool} --- the dynamical mass measurement is now in agreement with the spectrum-derived temperature and independently measured age. 

In this work, we presented follow-up spectroscopic imaging on recently-discovered directly imaged brown dwarf companion HIP 93398 B using Subaru SCExAO/CHARIS in broadband (\textit{J}/\textit{H}/\textit{K}), \textit{H} band, and \textit{J} band. Using a new epoch of data from 2023 combined with existing relative astrometry, HARPS RVs, and \textit{Hipparcos}/\textit{Gaia} absolute astrometry, we support previous measurements of the object's dynamical mass of $\sim$66 $M_{\rm Jup}$, establishing this object as an old ($3.59^{+0.87}_{-1.15}$ Gyr) and massive brown dwarf companion. Spectral analysis reveals this object to be a cloudy \revv{late-L brown dwarf near the L/T transition}, contrary to previous classifications as a T dwarf. \revv{With the revised} temperature and luminosity in evolutionary models (e.g.~Sonora Diamondback), HIP 93398 B is no longer in tension with evolutionary predictions; although there are a handful of such T class objects that are in disagreement with evolutionary models (e.g. Gl 229 B, \citet{calamari2022atmospheric}), HIP 93398 B does not seem to be among them. 

Comparing so-called \revv{``overmassive T dwarfs''} with new generations of substellar atmosphere models featuring clouds and disequilibrium chemistry may \revv{affect derived temperatures and accordingly} resolve some of the disagreement between models and observations. The growing sample of brown dwarfs with dynamical mass measurements, identified by HGCA-informed surveys like HIP 93398 B, will be key for continued testing of substellar atmosphere models.

\clearpage
\section*{Acknowledgments}
Thank you to Mark Popinchalk, Emily Calamari, and Genaro Suarez for providing helpful resources on brown dwarf atmospheres to BLL, and to Kellen Lawson and Thayne Currie for advice on up-to-date CHARIS calibrations.

This material is based upon work supported by the National Science Foundation Graduate Research Fellowship under Grant No. 2021-25 DGE-2034835 for author BLL. Any opinions, findings, and conclusions or recommendations expressed in this material are those of the authors(s) and do not necessarily reflect the views of the National Science Foundation. TDB acknowledges support from the Alfred P. Sloan Foundation and from the NASA Exoplanet Research Program under grant \#80NSSC18K0439.

This work presents results from the European Space Agency (ESA) space mission \textit{Gaia}. \textit{Gaia} data are being processed by the \textit{Gaia} Data Processing and Analysis Consortium (DPAC). Funding for the DPAC is provided by national institutions, in particular the institutions participating in the \textit{Gaia} MultiLateral Agreement (MLA). The \textit{Gaia} mission website is https://www.cosmos.esa.int/gaia. The \textit{Gaia} archive website is https://archives.esac.esa.int/gaia. This research has benefitted from the SpeX Prism Spectral Libraries, maintained by Adam Burgasser at http://www.browndwarfs.org/spexprism.

This work has benefitted from The UltracoolSheet at http://bit.ly/UltracoolSheet, maintained by Will Best, Trent Dupuy, Michael Liu, Aniket Sanghi, Rob Siverd, and Zhoujian Zhang, and developed from compilations by \citet{dupuy2012hawaii}, \citet{dupuy2013distances}, \citet{deacon2014wide}, \citet{liu2016hawaii}, \citet{best2017photometry}, \citet{best2020volume}, \citet{sanghi2023hawaii}, and \citet{schneider2023astrometry}.		

This research has made use of the Keck Observatory Archive (KOA), which is operated by the W. M. Keck Observatory and the NASA Exoplanet Science Institute (NExScI), under contract with the National Aeronautics and Space Administration. B. Lewis would also like to acknowledge the invaluable labor of the maintenance and clerical staff at her institution and the observatories, whose contributions make scientific discovery a reality. This research was partially conducted on the traditional lands of the Gabrielino-Tongva people. 

This research is based on data collected at the Subaru Telescope, which is operated by the National Astronomical Observatory of Japan.  The development of SCExAO is supported by the Japan Society for the Promotion of Science (Grant-in-Aid for Research \#23340051, \#26220704, \#23103002, \#19H00703, \#19H00695 and \#21H04998), the Subaru Telescope, the National Astronomical Observatory of Japan, the Astrobiology Center of the National Institutes of Natural Sciences, Japan, the Mt Cuba Foundation and the Heising-Simons Foundation. CHARIS was built at Princeton University under a Grant-in-Aid for Scientific Research on Innovative Areas from MEXT of the Japanese government (\#23103002). We are honored and grateful for the opportunity of observing the Universe from Maunakea, which has cultural, historical and natural significance in Hawai'i.

\clearpage
%


\software{NumPy \citep{numpy},
IPython \citep{ipython}, 
Jupyter Notebooks \citep{jupyter},
Matplotlib \citep{matplotlib}, 
Astropy \citep{astropy:2013, astropy:2018}, 
SciPy \citep{scipy}, emcee \citep{Foreman-Mackey2013PASP}, pyklip-CHARIS \citep{Wang2015ascl,Chen2023RASTI}, pysynphot \citep{STScIDevelopmentTeam2013ascl}, SPLAT \citep{Burgasser2017ASInC}, orvara \citep{brandt2021orvara}, orbitize! \citep{2020AJ....159...89B}, ptemcee \citep{2016MNRAS.455.1919V}}

\vspace{5mm}




\bibliography{charis}
\bibliographystyle{aasjournal}



\end{document}